%% file: NOPG.tex
\documentclass[12pt,titlepage,acknowledgements,a4paper]{article}
\usepackage{amsmath,amsfonts,amsthm,amssymb,euscript,eufrak,mathrsfs,epsf,epsfig}
\usepackage{array,verbatim}
\usepackage[sort&compress]{natbib}
\makeatletter

\makeatletter
\@addtoreset{equation}{section}
\makeatother

\newcounter{multieqs}

\input{DansDefs}

\begin{document}

\begin{flushright}
QMUL-PH-10-20
\end{flushright}

\vspace{50pt}

\begin{center}

{\Large \bf Comments on Higher Order BLG Supersymmetry Transformations}\\
\vspace{33pt}

{\bf {\mbox{Andrew M. Low}}}%
\footnote{{{\tt a.m.low}{\tt @qmul.ac.uk }}}

{\em Centre for Research in String Theory \\ Department of Physics\\
Queen Mary, University of
London\\
Mile End Road, London, E1 4NS\\
United Kingdom}
\vspace{40pt}

{\bf Abstract}

\end{center}

\noindent
In this paper we begin an investigation into the $l_p^3$ corrections to the supersymmetry transformations of the Bagger-Lambert-Gustavsson (BLG) theory. We begin with a review of the dNS duality transformation which allows a non-abelian gauge field to be dualised to a scalar field in 2+1 dimensions. Applying this duality to $\alpha'^2$ terms of the non-abelian D2-brane theory gives rise to the $l_p^3$ corrections of the Lorentzian BLG theory. We then apply this duality transformation to the ${\alpha'^2}$ corrections of the D2-brane supersymmetry transformations. For the `abelian' BLG theory we are able to uniquely determine the $l_p^3$ corrections to the supersymmetry transformations of the scalar and fermion fields. Generalising to the `non-abelian' BLG theory we are able to determine the $l_p^3$ correction to the supersymmetry transformation of the fermion field. Along the way make a number of observations relating to the implementation of the dNS duality transformation at the level of supersymmetry transformations.
\vspace{0.5cm}

\setcounter{page}{0}
\thispagestyle{empty}
\newpage

\section{Introduction}
The BLG Lagrangian and supersymmetry transformations presented in \cite{Gustavsson:2007vu, Bagger:2006sk, Bagger:2007jr} can be thought of as representing the leading order terms in an $l_p$ expansion of a (not yet determined) non-linear M2-brane theory. This is analogous to the fact that super Yang-Mills theory represents the leading order terms of the non-abelian Born-Infeld action, which is believed to describe the dynamics of coincident D-branes.\footnote{Note that the symmetrised trace prescription of the non-abelian Born-Infeld action \cite{Tseytlin:1997csa}  breaks down at sixth order and higher in the worldvolume field strength \cite{Hashimoto:1997gm}.}  Ultimately one would like to determine the full theory, of which the leading order terms are those of the BLG Lagrangian. Toward this end it is constructive to consider the next order in $l_p$ corrections to the theory. At the level of the Lagrangian this analysis has been performed in the literature \cite{Alishahiha:2008rs, Ezhuthachan:2009sr} using two complimentary methods.\footnote{For other discussions on non-linear corrections to Bagger-Lambert theory see \cite{Kluson:2008nw, Iengo:2008cq, Sasaki:2009ij, Li:2008ya, Garousi:2008xn}.}

The first method involves a duality transformation due to de-Witt, Nicholai and Samtleben (dNS). This duality is based on the fact that in (2+1) dimensions, a gauge field is dual to a scalar, and it is therefore possible to replace the gauge-field with a scalar field such that the theory possesses a manifest SO(8), rather than SO(7) symmetry. In \cite{Ezhuthachan:2008ch}, it was shown that applying this procedure to the D2-brane Lagrangian, it is possible to re-write the theory as a Lorentzian Bagger-Lambert theory. This technique was then applied to the ${\alpha'}^2$ terms of the D2-brane Lagrangian in order to determine the $l_p^3$ corrections to the Lorentzian BLG theory\footnote{Lorentzian Bagger-Lambert theories are considered in \cite{Gomis:2008uv, Ho:2008ei, Benvenuti:2008bt, Bandres:2008kj, Gomis:2008be}. See also \cite{Honma:2008un, Honma:2008ef, Honma:2008jd, Antonyan:2008jf}.}. Remarkably, all higher order Lagrangian terms were expressible in terms of basic building blocks involving covariant derivatives, $D_\mu X^I$ and three-brackets $[X^I, X^J, X^K]$. This led the authors of \cite{Alishahiha:2008rs} to conjecture that the higher derivative Lagrangian they had derived would also apply to the ${\mathcal{A}}_4$ BLG Theory. This conjecture was confirmed in \cite{Ezhuthachan:2009sr} where the novel Higgs mechanism \cite{Mukhi:2008ux} was used to determine the ${\mathcal{A}}_4$ theory Lagrangian at order $l_p^3$. This involved using dimensional analysis to write down all possible $l_p^3$ corrections to the BLG Lagrangian with arbitrary coefficients. The coefficients were fixed by applying the novel Higgs mechanism to the higher order terms and matching them to the ${\alpha'}^2$ terms of the D2-brane theory. It was shown that the structure of the higher order terms in both the ${\mathcal{A}}_4$ and Lorentzian theories take the same form. 

Given that the $l_p^3$ corrections to the BLG theory have been calculated, one might ask whether these terms are maximally supersymmetric, and if so, to determine the structure of the higher order supersymmetry transformations. In this paper we begin the task of calculating the $l_p^3$ corrections to the supersymmetry transformations of the BLG theory.
The hope is that closure of the higher order supersymmetry transformations would uniquely determine the higher order corrections to the BLG equations of motion which can then be `integrated' to determine the higher order Lagrangian, which by definition, would be supersymmetric. One could in principle write down all possible $l_p^3$ corrections to the supersymmetry transformations and then try and fix the coefficients by demanding the closure of the supersymmetry algebra. However the plethora of possible terms at order $l_p^3$ would make the closure of the algebra a mammoth task. To try and simplify the problem we will use the non-abelian D2-brane theory as a guide. We know that dNS duality transformation allows us to map the non-abelian D2-brane Lagrangian into the Lorentzian BLG Lagrangian. Furtherore we know that the structure of this Lagrangian is the same as the structure of the ${\mathcal{A}}_4$ theory Lagrangian. It is natural to ask whether this methodology can tell us anything about how the higher order D2-brane supersymmetry transformations are related to the $l_p^3$ corrections to the BLG supersymmetry transformations. 

In the first part of this paper we will review the dNS duality transformation \cite{Nicolai:2003bp, deWit:2003ja, deWit:2004yr} by considering how the Lorentzian BLG Lagrangian can be derived from the D2-brane theory. We will then attempt to apply the duality transformation at the level of supersymmetry transformations. To simplify the task we will begin by only considering the `abelian' BLG theory. We will see that the duality transformation works for the fermion variation but fails to work for the scalar variation. Therefore in order to calculate the scalar variation we have to use a different approach. This involves using dimensional analysis to write the most general scalar variation with arbitrary coefficients. Invariance of the higher order Lagrangian is then used to fix the values of the coefficients. 
In the final part of this paper we begin the task of calculating the full `non-abelian' BLG supersymmetry transformations at $\mathcal{O} (l_p^3)$. We are able to uniquely determine the higher order fermion variation but unable to uniquely determine the scalar variation. As a result, this paper represents work in progress.
\section{Non-abelian duality in 2+1 dimensions}
We begin by reviewing a prescription for dualising non-abelian gauge fields in (2+1) dimensions due to de Wit, Nicolai and Samtleben (dNS)\cite{Nicolai:2003bp, deWit:2003ja, deWit:2004yr}. We will follow the presentation of \cite{Ezhuthachan:2008ch}. According to the dNS prescription the Yang-Mills gauge field $A_\mu$ gets replaced by two non-dynamical gauge fields $A_\mu$ and $B_\mu$ with a $B \wedge F$ type kinetic term, plus an extra scalar which ends up carrying the dynamical degrees of freedom of the original Yang-Mills gauge field. The duality transformation is enforced by making the replacement
\begin{equation}
\Tr \left(-\frac{1}{4 g^2_{YM}} F_{\mu \nu} F^{\mu \nu} \right) \rightarrow \Tr \left(\frac{1}{2} \epsilon^{\mu \nu \lambda}B_\mu F_{\nu \lambda} - \frac{1}{2}(D_\mu \phi-g_{YM}B_\mu)^2 \right). \label{kunt}
\end{equation}
We wish to consider the effect of this transformation on the multiple D2-brane theory. The low energy Lagrangian for this theory is obtained by reducing ten-dimensional $U(N)$ super Yang-Mills theory to (2+1) dimensions. In this case, making the replacement \eqref{kunt} in the D2-brane Lagrangian results in the dNS transformed Lagrangian\footnote{This action exhibits an abelian gauge symmetry  allowing one to pick a gauge in which either $D^\mu B_\mu = 0$ or $\phi = 0$. In the latter case $B_\mu$ becomes an auxiliary field that can be integrated out thereby showing the equivalence of the LHS and RHS of \eqref{kunt}. For explicit details see \cite{Ezhuthachan:2008ch}. } 
\begin{align}
\mathcal{L}= &\Tr ( \frac{1}{2} \epsilon^{\mu \nu \lambda}B_\mu F_{\nu \lambda} - \frac{1}{2}(D_\mu \phi-g_{YM}B_\mu )^2- \frac{1}{2} D_\mu X^i D^\mu X^i \nn \\
&- \frac{g^2_{YM}}{4}[X^i, X^j][X^i, X^j]  + \frac{i}{2} \bp \g^\mu D_\mu \psi + \frac{i}{2} g_{YM} \bp \g_i [X^i, \psi ] ).
\end{align}
The gauge invariant kinetic terms for the eight scalars can be shown to possess an SO(8) invariance by renaming $\phi \rightarrow X^8$ and writing
\begin{align}
{\tilde{D}}^\mu X^i &= D_\mu X^i = \partial_\mu X^i - [A_\mu, X^i], \quad i=1,2, \ldots, 7 \\
 {\tilde{D}}^\mu X^8 &= D_\mu X^8 - g_{YM}B_\mu = \partial_\mu X^8 - [A_\mu, X^8] - g_{YM}B_\mu.
\end{align}
Defining the constant 8-vector
\begin{equation}
g^I_{YM} = (0, \ldots, 0, g_{YM}), \quad I=1,2, \ldots, 8,
\end{equation}
allows one to define the covariant derivative
\begin{equation}
{\tilde{D}}^\mu X^I = D_\mu X^I - g^I_{YM} B_\mu.
\end{equation}
It is then possible write the super Yang-Mills action in a form that is $SO(8)$ invariant under transformations that rotate both the fields $X^I$ and the coupling constant vector $g^I_{YM}$:
\begin{align}
\mathcal{L} = \Tr \left( \frac{1}{2} \e^{\mu \nu \lambda} B_\mu F_{\nu \lambda} - \frac{1}{2} ({\tilde{D}}_\mu X^I)^2 + \frac{i}{2} \bp \g^\mu D_\mu \psi + \frac{i}{2} g^I_{YM} \bp \g_{IJ} [X^J , \psi ] -\frac{1}{12} ({X^{IJK}})^2 \right)
\end{align}
where the three-bracket $X^{IJK}$ is defined as
\begin{equation}
X^{IJK} = g^I_{YM} [X^J, X^K] + g^J_{YM} [X^K, X^I] + g_{YM}^K [X^I , X^J].
\end{equation}
This theory is only formally SO(8) invariant, as the transformations must act on the coupling constants as well as the fields. However, one can replace the vector of coupling constants $g^I_{YM}$ by a new (gauge singlet) scalar $X^I_+$ provided that the new scalar field has an equation of motion that renders it constant.  Constancy of $X^I_+$ is imposed by adding a new term to the Lagrangian involving a set of abelian gauge fields and scalars $C_\mu^I$ and $X_-^I$: 
\begin{equation}
{\mathcal{L}}_C = (C^\mu_I - \partial^\mu X^I_-) \partial_\mu X^I_+.
\end{equation}
As explained in \cite{Ezhuthachan:2008ch, Gomis:2008be}, this term has the effect of constraining the vector $X_+^I$ to be an arbitrary constant which can be identified with $g^I_{YM}$. In this way one recovers the gauge-fixed Lorentzian models of \cite{Bandres:2008kj, Gomis:2008be}. One might wonder whether this non-abelian duality works when higher order (in $\alpha'$) corrections are included in the D2-brane theory. In particular, does the 3-algebra structure survive $\alpha'$ corrections? In \cite{Alishahiha:2008rs, Ezhuthachan:2009sr} it was shown that at $\mathcal{O} ({\alpha'}^2)$ the duality does work and all terms in the resulting $l_p^3$ corrected M2-brane theory are expressible in terms of ${\tilde{D}}_\mu X^I$ and $X^{IJK}$ building blocks. Another question one might ask is whether this duality works at the level of supersymmetry transformations and if so, would it be possible to derive the $\mathcal{O} (l_p^3)$ corrections to the BLG supersymmetry transformations? The first step towards answering this question is to consider \textit{how} abelian duality in (2+1) dimensions can be implemented at the level of supersymmetry transformations. To this we now turn.

\section{Abelian Duality and Supersymmetry}
Our ultimate objective is to determine higher order supersymmetry transformations in BLG theory by using the dNS procedure outlined in the previous section. As a warm-up exercise we will consider dualising abelian gauge-fields to scalars in 2+1 dimensions and see how this works at the level of supersymmetry transformations for a single D2-brane. Let us begin by considering abelian duality at the level of the Lagrangian.
\subsection{Abelian Duality}
Consider the following 2+1 dimensional action involving a Lagrange multiplier field $X$
\begin{equation}
S = - \int d^3 \sigma (\frac{1}{4} F_{\mu \nu} F^{\mu \nu} + \frac{1}{2} \e_{\mu \nu \lambda} F^{\mu \nu} \partial^\lambda X )\label{lagrange}
\end{equation}
We see that the gauge field equation of motion takes the form
\begin{equation}
F_{\mu \nu} = - \e_{\mu \nu \lambda} \partial^\lambda X \label{gf}
\end{equation}
whereas the $X$ equation of motion takes the form of the Bianchi identity
\begin{equation}
\e_{\mu \nu \lambda} \partial^\mu F^{\nu \lambda} = 0.
\end{equation}
If we substitute the gauge field equation of motion into \eqref{lagrange} then we find a kinetic term for $X$
\begin{equation}
\int d^3 \sigma (\frac{1}{4} F_{\mu \nu} F^{\mu \nu} + \frac{1}{2} \e_{\mu \nu \lambda} F^{\mu \nu} \partial^\lambda X ) \rightarrow  \int d^3 \sigma \frac{1}{2}\partial_\mu X \partial^\mu X.
\end{equation} 
Alternatively, use of the Bianchi identity in \eqref{lagrange} results in 
\begin{equation}
\int d^3 \sigma (\frac{1}{4} F_{\mu \nu} F^{\mu \nu} + \frac{1}{2} \e_{\mu \nu \lambda} F^{\mu \nu} \partial^\lambda X ) \rightarrow  \int d^3 \frac{1}{4} F_{\mu \nu} F^{\mu \nu}. 
\end{equation} 
with $F = dA$. So how does this relate to the D2-brane theory? The leading order Lagrangian for a single D2-brane can be obtained by dimensional reduction of super Yang-Mills theory in ten dimensions. The bosonic D2-brane action can be expressed as
\begin{equation}
S = \int d^3\sigma (-\frac{1}{4}F_{\mu \nu} F^{\mu \nu} - \frac{1}{2} \partial_\mu X^i \partial^\mu X^i). \label{vjy}
\end{equation}
Abelian duality is implemented by making the replacement 
\begin{equation}
-\frac{1}{4} F_{\mu \nu} F^{\mu \nu} \rightarrow - (\frac{1}{4} F_{\mu \nu} F^{\mu \nu} + \frac{1}{2} \e_{\mu \nu \lambda} \partial^\mu X F^{\nu \lambda})
\end{equation}
in the action \eqref{vjy}. Use of the gauge field equation of motion \eqref{gf} then results in
\begin{align}
S &= \int d^3\sigma (-\frac{1}{4}F_{\mu \nu} F^{\mu \nu} - \frac{1}{2} \e_{\mu \nu \lambda} F^{\mu \nu} \partial^\lambda X - \frac{1}{2} \partial_\mu X^i \partial^\mu X^i) \nn \\
&= \int d^3 \sigma ( -\frac{1}{2} \partial_\mu X \partial^\mu X - \frac{1}{2} \partial_\mu X^i \partial^\mu X^i) \nn \\
&= \int d^3 \sigma (- \frac{1}{2} \partial_\mu X^I \partial^\mu X^I)
\end{align}
where in obtaining the last line we identified $X = X^8$ as the eighth scalar field. We see that the scalar kinetic term now has the desired $SO(8)$ invariant form. Note that it is possible to implement abelian duality in (2+1) dimensions at the level of the full DBI action \cite{Bergshoeff:1996tu}. In this way one is able to derive a non-linear Lagrangian for a membrane in the static gauge with the expected SO(8) symmetry. Now that we have seen how abelian duality works at the level of the Lagrangian let us consider applying this to the supersymmetry transformations of a single D2-brane.
\subsection{supersymmetry transformations}
The D2-brane supersymmetry transformation can be obtained by dimensionally reducing the supersymmetry transformations of ten dimensional super Yang-Mills. The spinors appearing in the 10-dimensional theory are Majorana-Weyl and satisfy
\begin{equation}
\g^{(10)} \chi = \chi
\end{equation}
where $\g^{(10)}$ is the ten dimensional chirality matrix. Since we are interested in uplifting the D2-brane theory to M-theory it is desirable to look for an embedding of $SO(1,9)$ into $SO(1,10)$ in which $\g^{(10)}$ becomes the eleventh gamma matrix. We denote the gamma matrices of $SO(1,10)$ as $\g^M (M = 0, \ldots , 9, 10 )$. In eleven dimensions the spinors will be Majorana. However we know that the presence of the M2-brane breaks the Lorentz symmetry as $SO(1,10) \rightarrow SO(1,2) \times SO(8)$ and therefore we can have a Weyl spinor of $SO(8)$. Let us denote the chirality matrix of $SO(8)$ by $\g$ where
\begin{equation}
\g = \g^{3 \ldots 9(10)}
\end{equation}
The M2-brane breaks half the supersymmetry of the vacuum. We choose conventions in which 
\begin{equation}
\g \e = \e, \qquad \g \psi = - \psi.
\end{equation}
Under dimensional reduction the (9+1) dimensional gauge field will split into a (2+1)-dimensional gauge field $A_\mu$ and a scalar field $X^i$ transforming under $SO(7)$. As usual with dimensional reduction, the fields are independent of the circle directions such that one can set $\partial_i = 0$. In what follows, for reasons that will become clear shortly, we will label $\mu = 0,1,2$ and $i = 1, \ldots 7$ with the ten-dimensional chirality matrix relabeled as $\g^{(10)} = \g^8$. Dimensional reduction of the ten-dimensional super Yang-Mills transformations
\begin{align}
\delta A^M &= i\be \g^M \psi \nn \\
\delta \psi &= \frac{1}{2} \g^{MN} F_{MN} \e
\end{align}
results in the following D2-brane transformations 
\begin{align}
\delta X^i &= i \be \g^i \psi  \\
\delta A_\mu &= i \be \g_\mu \g^8 \psi  \label{one} \\
\delta \psi &= \frac{1}{2} \g^{\mu \nu} F_{\mu \nu} \g^8\e + \g^\mu \g^i \partial_\mu X^i \e \label{too}
\end{align}
We now consider the effect of applying abelian duality at the level of supersymmetry transformations. This can be achieved by using \eqref{gf} to write
\begin{equation}
\partial_\mu X^8 = \frac{1}{2} \e_{\mu \nu \lambda} F^{\nu \lambda} \label{pohg}
\end{equation}
where we have relabeled the scalar appearing in \eqref{gf} as $X^8$ (this will provide the `eighth' scalar which will combine with the other seven to give an SO(8) invariant supersymmetry transformation). Performing the duality transformation on the fermion variation involves substituting \eqref{pohg} into \eqref{too} 
\begin{align}
\delta \psi &= \frac{1}{2} \e_{\mu \nu \lambda} \g^{\mu \nu} \partial^\lambda X^8 \g^8 \e + \g^\mu \g^i \partial_\mu X^i \e \nn \\
&= \g^\mu \g^8 \partial_\mu X^8 \e + \g^\mu \g^i \partial_\mu X^i \e \nn \\
&= \g^\mu \g^I \partial_\mu X^I \e.
\end{align}
We see that this now takes the desired SO(8) form.  In order to determine the SO(8) transformation of the scalar field $\delta X^I$ we need to consider \eqref{one} re-written as
\begin{equation}
\delta F_{\mu \nu} = - 2i \be \g_{[ \mu} \g^8 \partial_{\nu ]} \psi.
\end{equation}
Substituting \eqref{pohg} into the left-hand side of this transformation allows us to write
\begin{align}
 \partial^\lambda \delta X^8 &= - i \be \e^{\mu \nu \lambda} \g_{[\mu} \g^8 \partial_{\nu ]} \psi \nn \\
&= i \be \g^{\nu \lambda } \g^8 \partial_\nu \psi \nn \\
&= i \be (\eta^{\nu \lambda} - \g^\lambda \g^{\nu}) \g^8 \partial_\nu \psi \nn \\
&= i \be \g^8 \partial^\lambda \psi \label{bt}
\end{align}
where we have made use of the lowest order fermion equation of motion $\g^\mu \partial_\mu \psi = 0$. This relation implies that $\delta X^8 = i \be \g^8 \psi$ which can be combined with $\delta X^i = i \be \g^i \psi$ to give
\begin{equation}
\delta X^I = \be \g^I \psi.
\end{equation}
In summary we see that at lowest order it is possible to re-write the D2-brane supersymmetry transformations in an SO(8) invariant form. For the fermion variation this simply involved substituting \eqref{pohg} into the D2-brane expression. For the scalar field variation it was necessary to `dualise' the gauge field variation $\delta F_{\mu \nu}$ to form the eighth scalar. In the next section we extend our analysis to higher order abelian supersymmetry transformations. 

\section{Higher Order Abelian Supersymmetry }
In this section we will determine the $l_p^3$ corrections to the abelian M2-brane supersymmetry transformations (excluding bi-linear and tri-linear fermion terms). We begin by using dimensional arguments to determine the structure of the supersymmetry transformations. We will then apply the duality transformation outlined in the previous section to the $\mathcal{O}({\alpha'}^2)$ D2-brane supersymmetry transformations. We will see that this procedure uniquely determines the fermion variation but fails to work for the scalar variation. This will motivate us to try a different approach.   
\subsection{Dimensional Analysis}
Dimensional analysis tells us that the mass dimensions of the fields appearing in the BLG theory are
$ [X] = \frac{1}{2} , [\psi ] = [ A_\mu ] = 1$. The supersymmetry parameter $\e$ carries mass dimension $[\e] = -\frac{1}{2}$. We expect the first non-trivial corrections to the supersymmetry transformations to appear at $\mathcal{O} (l_p^3)$. Therefore we see that the $\mathcal{O} (l_p^3)$ terms in $\delta \psi$ must be mass dimension $4$. In a similar manner the correction terms in $\delta X$ must be mass dimension $3 \frac{1}{2}$. For the sake of simplicity we will neglect bi-linear fermion terms in the scalar variation and tri-linear fermion terms in the fermion variation. In terms of the basic building blocks of scalar fields and derivatives, the only possible types of term appearing in the fermion variation at $\mathcal{O} (l_p^3)$ are those involving three derivatives and three scalar fields. If we assume that derivatives must always act on scalars (with at most one derivative) then a little thought reveals that the higher-order fermion variation takes the form
\begin{align}
\delta \psi = &+ a_1 l_p^3\g_\mu \g^I \partial_\nu X^J \partial^\nu X^J \partial^\mu X^I \e + a_2l_p^3 \g_\mu \g^I \partial^\mu X^J \partial_\nu X^J \partial^\nu X^I \e \nn \\
&+ a_3l_p^3\e^{\mu \nu \rho} \g^{IJK} \partial_\mu X^I \partial_\nu X^J \partial_\rho X^K \e. \label{fermi}
\end{align}
The motivation for assuming that scalars are always acted on by derivatives is based on the form of the ${\alpha'}^2$ D2-brane supersymmetry transformations (derived in the next section) which have no free scalar terms. Let us now consider the scalar transformation. Based on dimensional analysis and the reasons already outlined, the only types of term appearing in the scalar variation are those involving two derivatives, two scalar fields and a fermion. Considering all independent index contractions one arrives at the following expression
\begin{align}
\delta X^I =  &+ b_1 l_p^3\be \g^I \psi \partial_\mu X^J \partial^\mu X^J + b_2 l_p^3 \be \g^J \psi \partial_\mu X^I \partial^\mu X^J  \nn \\
&+ b_3 l_p^3\be \g^J \g^{\mu \nu} \psi \partial_\mu X^I \partial_\nu X^J + b_4 l_p^3\be \g^{\mu \nu} \g^{IJK} \psi \partial_\mu X^J \partial_\nu X^K. \label{bose}
\end{align}
Our task in the remainder of this section is to fix the coefficients appearing in \eqref{fermi} and \eqref{bose}. There are a number of ways that this can be achieved. In the next section we will attempt to fix these coefficients through abelian duality. We will see that this only works for the fermion variation. In order to determine the scalar variation we will have to use a different approach. This will involve checking that the higher order abelian Lagrangian derived in \cite{Ezhuthachan:2009sr} is invariant under \eqref{fermi} and \eqref{bose}. Not only will this allow us to determine the scalar variation coefficients but it will also provide a test for the fermion terms derived using the duality approach. 

\subsection{Higher Order Abelian Supersymmetry via Dualisation}
In this section we will attempt to derive the higher order abelian supersymmetry transformations using abelian duality. Our starting point will be the $\mathcal{O} (\alpha'^2)$ supersymmetry transformations of the ten-dimensional super Yang-Mills theory. These were first discovered by Bergshoeff and collaborators in \cite{Bergshoeff:1986jm}\footnote{The form of these transformations was later confirmed in \cite{Bergshoeff:2001dc, Cederwall:2001bt, Cederwall:2001td}.}
\begin{align}
\delta \psi &= \alpha'^2 (\lambda_1 \g^{MN} F_{PQ} F^{PQ} F_{MN} \e + \lambda_2 \g^{MN} F_{MP} F^{PQ} F_{QN} \e + \lambda_3 \g^{MNPQR}F_{MN}F_{PQ}F_{RS} \e). \nn \\
\delta A_M &= \alpha'^2 ( \alpha_1 \be \g_M  F_{NP} F^{NP} \psi + \alpha_2 \be \g_N  F_{MP}F^{PN} \psi + \alpha_3\be \g^{NPQ} F_{MN}F_{PQ} \psi \nn \\
&\quad + \alpha_4\be \g_{MNPQR}  F^{NP}F^{QR} \psi ).\label{ymtran}
\end{align}
Note that in \cite{Bergshoeff:1986jm} the fermion variation also included tri-linear fermion terms and the gauge field variation included bi-linear fermion terms which we have not included for the sake of simplicity. We have purposely left the coefficients unspecified. The hope is that these coefficients will be fixed by the requirement that the (2+1) dimensional transformations collect into $SO(8)$ invariant terms under the duality transformation. Next we wish to reduce these expressions to (2+1) dimensions. We will first focus on the fermion. Performing the dimensional reduction one finds
\begin{align}
\lambda_1 \g^{MN} F_{PQ} F^{PQ} F_{MN} \e \rightarrow &+ \lambda_1 \g^{\mu \nu} F_{\mu \nu} F^{\rho \sigma} F_{\rho \sigma} \e + 2\lambda_1 \g^{\mu \nu} F_{\mu \nu} \partial^\rho X^i \partial_\rho X^i \e \nn \\
&+ 2 \lambda_1 \g^\mu \g^i \partial_{\mu} X^i F_{\rho \sigma} F^{\rho \sigma} \e + 4 \lambda_1 \g^{\mu} \g^i \partial_\mu X^i \partial^\nu X^j \partial_\nu X^j \e.  \nn \\
\end{align}
\begin{align}
\lambda_2 \g^{MN} F_{MP} F^{PQ} F_{QN} \e \rightarrow &+ \lambda_2 \g^{\mu \nu} F_{\mu \rho} F^{\rho \sigma} F_{\sigma \nu} \e - 2\lambda_2 \g^{\mu \nu} F_{\mu \rho} \partial^\rho X^i \partial_\nu X^i \e  \nn \\
&+ 2\lambda_2 \g^\mu \g^i F_{\mu \rho} F^{\rho \sigma} \partial_\sigma X^i \e \nn \\
&- 2 \lambda_2\g^\mu \g^i \partial_\mu X^j \partial^\rho X^j \partial_\rho X^i \e  - \lambda_2 \g^{ij} \partial_\rho X^i F^{\rho \sigma} \partial_\sigma X^j \e. \nn 
\end{align}
\begin{align}
\lambda_3 \g^{MNPQRS} F_{MN} F_{PQ} F_{RS} \e \rightarrow &- 8 \lambda_3 \g^{\mu \nu \rho} \g^{ijk} \partial_\mu X^i \partial_\nu X^j \partial_\rho X^k \e. 
\end{align}
Next we dualise the gauge field using \eqref{pohg}. After a small amount of algebra one finds
\begin{align}
\lambda_1\g^{MN} F_{PQ} F^{PQ} F_{MN} \e \rightarrow &+4\lambda_1 \g^\mu \partial_\mu X^8 \partial^\nu X^8 \partial_\nu X^8\e - 4\lambda_1 \g^\mu \partial_\mu X^8 \partial^\nu X^i \partial_\nu X^i\e  \nn \\
&- 4 \lambda_1\g^\mu \g^i \partial_\mu X^i \partial^\nu X^8 \partial_\nu X^8 \e + 4 \lambda_1\g^{\mu} \g^i \partial_\mu X^i \partial^\nu X^j \partial_\nu X^j \e. \label{dad}
\end{align}
\begin{align}
\lambda_2 \g^{MN} F_{MP} F^{PQ} F_{QN} \e \rightarrow &-2 \lambda_2 \g^\mu \partial_\nu X^8 \partial^\nu X^8 \partial_\mu X^8\e - 2 \lambda_2 \g^\mu \partial^\nu X^8 \partial_\mu X^i \partial^\nu X^i\e \nn \\
&+ 2 \lambda_2 \g^\mu \partial_\mu X^8 \partial^\nu X^i \partial_\nu X^i \e + 2 \lambda_2 \g^\mu \g^i \partial_\nu X^8 \partial^\nu X^8 \partial_\mu X^i \e \nn \\
&-2 \lambda_2 \g^\mu \g^i \partial^\nu X^8 \partial_\mu X^8 \partial^\nu X^i \e - 2 \lambda_2 \g^\mu \g^i \partial_\mu X^j \partial^\rho X^j \partial_\rho X^i \e \nn \\
&- \lambda_2 \g^{ij} \e^{\rho \sigma \lambda} \partial_\rho X^i \partial_\lambda X^8 \partial_\sigma X^j \e.
\end{align}
\begin{align}
\lambda_3 \g^{MNPQRS} F_{MN} F_{PQ} F_{RS} \e &\rightarrow - 8 \lambda_3 \g^{\mu \nu \rho} \g^{ijk} \partial_\mu X^i \partial_\nu X^j \partial_\rho X^k \e. \label{mum}
\end{align}
We would like to re-write these transformed expressions in terms of $SO(8)$ objects. The only possible $SO(8)$ objects involving three derivatives are those contained in \eqref{fermi}. The hope is that the fermion supersymmetry transformation should be expressible as a particular combination of these basic objects. More specifically, by noting that
\begin{align}
\g_\mu \g^I \partial_\nu X^J \partial^\nu X^J \partial^\mu X^I \e &\rightarrow  \g_\mu \g^i \partial_\nu X^j \partial^\nu X^j D^\mu X^i \e + \g_\mu \g^i \partial_\nu X^8 \partial^\nu X^8 \partial^\mu X^i \e \nn \\
& \quad + \g_\mu \g^8 \partial_\nu X^j \partial^\nu X^j \partial^\mu X^8 \e + \g_\mu \g^8 \partial_\nu X^8 \partial^\nu X^8 \partial^\mu X^8 \e \nn \\
\nn \\
\g_\mu \g^I \partial^\mu X^J \partial_\nu X^J \partial^\nu X^I &\rightarrow \g_\mu \g^i \partial^\mu X^j \partial_\nu X^j \partial^\nu X^i \e + \g_\mu \g^i \partial^\mu X^8 \partial_\nu X^8 \partial^\nu X^i \e \nn \\
& \quad + \g_\mu \g^8 \partial^\mu X^j \partial_\nu X^j \partial^\nu X^8 \e + \g_\mu \g^8 \partial^\mu X^8 \partial_\nu X^8 \partial^\nu X^8  \e 
\end{align}
we can write the terms in \eqref{dad}-\eqref{mum} as
\begin{align}
\delta \psi_{\textrm{abelian}} = &+4 \lambda_1 \g_\mu \g^I \partial_\nu X^J \partial^\nu X^J \partial^\mu X^I\e - 2 \lambda_2 \g_\mu \g^I  \partial^\mu X^J \partial_\nu X^J \partial^\nu X^I \e\nn \\
&+ (2 \lambda_2 - 8\lambda_1 ) (\g_\mu  \partial_\nu X^j \partial^\nu X^j \partial^\mu X^8 \e + \g_\mu \g^j  \partial_\nu X^8 \partial^\nu X^8 \partial^\mu X^j\e) \nn \\
&- 8 \lambda_3 \g^{\mu \nu \rho} \g^{ijk} \partial_\mu X^i \partial_\nu X^j \partial_\rho X^k \e - \lambda_2 \g^{ij} \e^{\rho \sigma \lambda} \partial_\rho X^i \partial_\lambda X^8 \partial_\sigma X^j \e. \label{magic}
\end{align}
The last line in \eqref{magic} can be expressed in $SO(8)$ form by noting   
\begin{equation}
\e^{\mu \nu \rho} \g^{IJK}  \partial_\mu X^I \partial_\nu X^J \partial_\rho X^K  \e \rightarrow \e^{\mu \nu \rho} \g^{ijk}   \partial_\mu X^i \partial_\nu X^j \partial_\rho X^k \e + 3 \e^{\mu \nu \rho} \g^{ij} \g^8 \partial_\mu X^i \partial_\nu X^j \partial_\rho X^8 \e.  \nn
\end{equation}
Provided with this information we see that it's possible to write \eqref{magic} in SO(8) form provided the coefficients are related as
\begin{equation}
\lambda_2 = 4 \lambda_1 ; \quad \lambda_2 = - 24 \lambda_3.
\end{equation}
The final result for the abelian fermion variation is
\begin{align}
\delta \psi = &+4 \lambda_1 l_p^3 \g_\mu \g^I \partial_\nu X^J \partial^\nu X^J \partial^\mu X^I\e - 8 \lambda_1 l_p^3\g_\mu \g^I  \partial^\mu X^J \partial_\nu X^J \partial^\nu X^I \e\nn \\
&-\frac{4}{3}\lambda_1 l_p^3\e^{\mu \nu \rho} \g^{IJK}  \partial_\mu X^I \partial_\nu X^J \partial_\rho X^K \e \label{magic2}.
\end{align}
A few comments are in order. Firstly we see that the structure of these terms exactly matches the structure of the terms appearing in \eqref{fermi} with the coefficients fixed as $a_1 = 4\lambda_1$, $a_2 = - 8\lambda_1$ and $a_3 = - 4/3 \lambda_1$. It remains to determine $\lambda_1$. Most remarkably we see that the requirement of SO(8) invariance has placed a constraint on the coefficients of the ten-dimensional supersymmetry transformations! Furthermore the ratios of the coefficients exactly matches the literature \cite{Collinucci:2002gd, Bergshoeff:1986jm, Cederwall:2001td, Collinucci:2002ac}. Thus it would appear that abelian duality does indeed work at the level of the fermion supersymmetry transformation. So what about the scalar variation? One would expect the higher order scalar supersymmetry transformation to work in a similar way to the lower order transformation. In other words, one expects the (2+1) dimensional gauge field transformation to contribute (after dualisation) to the `eighth' component of the scalar transformation $\delta X^I$. In order to see how this works we will need to determine $\delta F_{\mu \nu}$ in (2+1) dimensions. This can be constructed from our knowledge of $\delta A_\mu$. Therefore the first thing we need to do is dimensionally reduce the ten dimensional gauge field transformation $\delta A_M$ appearing in \eqref{ymtran}. Performing the reduction results in a scalar field supersymmetry transformation
\begin{align}
\delta X^i = &+\alpha_1 \be \g^i F_{\mu \nu} F^{\mu \nu} \psi + 2 \alpha_1 \be \g^i \partial_\mu X^j \partial^\mu X^j \nn \\
&+ \alpha_2 \be \g_\mu \partial_\rho X^i F^{\mu \rho} \psi - \alpha_2 \be \g^j \partial_\rho X^i \partial^\rho X^j \psi \nn \\
&- \alpha_3 \be \g^{\mu \nu \rho} \partial_\mu X^i F_{\nu \rho} \psi - \alpha_3 \be \g^{\mu \nu} \g^j \partial_\mu X^i \partial_\nu X^j \psi \nn \\
&- 4 \alpha_4 \be \g^{ijk} \g^{\mu \nu} \partial_\mu X^j \partial_\nu X^k \psi - 4 \alpha_4 \be \g^{ij} \g^{\mu \nu \rho} F_{\mu \nu} \partial_\rho X^j
\end{align}
and a gauge field supersymmetry transformation 
\begin{align}
\delta A_\mu = &+ \alpha_1 \be \g_\mu F_{\nu \rho} F^{\nu \rho} \psi + 2 \alpha_1 \be \g_\mu \partial_\rho X^i \partial^\rho X^i \psi \nn \\
&+ \alpha_2 \be \g_\rho F_{\mu \nu} F^{\nu \rho} \psi + \alpha_2 \be \g^i F_{\mu \nu} \partial^\nu X^i \psi \nn \\
&- \alpha_2 \be \g_\rho \partial^\rho X^i \partial_\mu X^i \psi + \alpha_3 \be \g^{\nu \rho \sigma} F_{\mu \nu} F_{\rho \sigma} \psi \nn \\
&+ 2 \alpha_3 \be \g^{\nu \rho} \g^i F_{\mu \nu} \partial_\rho X^i + \alpha_3 \be \g^{\rho \sigma} \g^i \partial_\mu X^i F_{\rho \sigma} \nn \\
&- 2 \alpha_3 \be \g^{ij} \g^\rho \partial_\mu X^i \partial_\rho X^j - 4 \alpha_4 \be \g_{\mu \nu \rho} \g^{ij} \partial^\nu X^i \partial^\rho X^j.
\end{align}
Performing the dualisation of the gauge field results in the scalar transformation 
\begin{align}
\delta X^i = &- 2 \alpha_1 \be \g^i \partial_\mu X^8 \partial^\mu X^8 \psi + 2 \alpha_1 \be \g^i \partial_\mu X^j \partial^\mu X^j \psi \nn \\
&+ \alpha_2 \be \g^{\mu \nu} \partial_\mu X^i \partial_\nu X^8 \psi - \alpha_2 \be \g^j \partial_\mu X^i \partial^\mu X^j \psi \nn \\
&- 2 \alpha_3 \be \partial_\mu X^i \partial^\mu X^8 \psi - 2 \alpha_3 \be \g^{\mu \nu} \g^j \partial_\mu X^i \partial_\nu X^j \psi \nn \\
&-8 \alpha_4 \be \g^{ij} \partial_\mu X^j \partial^\mu X^8 \psi - 4 \alpha_4 \be \g^{ijk} \g^{\mu \nu} \partial_\mu X^j
 \partial_\nu X^k \psi. \label{failure1}
\end{align}
Similarly for the gauge field one finds
\begin{align}
\delta A_\mu = &+ (\alpha_2 - 2 \alpha_1) \be \g_\mu \psi \partial_\nu X^8 \partial^\nu X^8  - \alpha_2 \be \g_\nu \psi \partial^\nu X^8 \partial_\mu X^8 \nn \\
&+ 2 \alpha_1 \be \g_\mu \psi \partial_\nu X^i \partial^\nu X^i - \alpha_2 \be \g_\nu \psi \partial_\mu X^i \partial^\nu X^i   \nn \\
&- 2\alpha_3 \be \g^{ij} \g^\sigma \psi \partial_\mu X^i \partial_\sigma X^j - 2 \alpha_3 \be \g_\mu \g^j \psi \partial^\nu X^8 \partial_\nu X^j \nn \\
 &+ \alpha_2 \e_{\mu \nu \lambda} \be \g^j \psi \partial^\lambda X^8 \partial^\nu X^j+ 4 \alpha_4 \e_{\mu \nu \rho} \be \g^{ij} \psi \partial^\nu X^i \partial^\rho X^j. \label{failure2}
\end{align}
The hope is that, just as for the fermion, these terms will combine into SO(8) invariant objects and in doing so fix the ratios of the coefficients. However we immediately encounter a problem which did not exist for the fermion variation. To see this let us focus on the first two terms appearing in \eqref{failure1}. These are the only two terms in \eqref{failure1} with the correct index structure to form the SO(8) term $\be \g^I \psi \partial_\mu X^J \partial^\mu X^J$ appearing in \eqref{bose}. However there is a relative minus sign appearing in these two terms meaning they are unable to combine. The problem can be traced back to the ten-dimensional gauge-field transformation term $\be \g_M  F_{NP} F^{NP} \psi$. 
Why is this happening? After all, we know that abelian duality works for the D2-brane Lagrangian and we know that the D2-brane Lagrangian derives from the ten-dimensional Yang-Mills term $-\frac{1}{4}F_{MN}F^{MN}$. Indeed, upon dimensional reduction of the ten-dimensional Yang-Mills Lagrangian and application of \eqref{pohg} one is left with a term $\frac{1}{2}\partial_\mu X^8 \partial^\mu X^8 -\frac{1}{2} \partial_\mu X^i \partial^\mu X^i$ which will not combine to form an SO(8) invariant scalar kinetic term. The way this problem is solved at Lagrangian level is by adding a Lagrange multiplier term $\frac{1}{2} \e_{\mu \nu \lambda} \partial^\mu X^8 F^{\nu \lambda}$ which under dualisation (according to \eqref{pohg}) combines with $\frac{1}{2}\partial_\mu X^8 \partial^\mu X^8$ in such a way as to change the sign of this term thereby allowing it to combine with $-\frac{1}{2} \partial_\mu X^i \partial^\mu X^i$ to form the desired SO(8) invariant scalar kinetic term. This suggests that the problem may in fact be the prescription \eqref{pohg}. In order to implement the duality on $F^2$ terms it may be necessary to make the replacement $\frac{1}{4}F^2 \rightarrow \frac{1}{4}F^2 + \frac{1}{2} \e_{\mu \nu \lambda} \partial^\mu X^8 F^{\nu \lambda}$. However, if this is true then it's unclear why making the replacement \eqref{pohg} works for the fermion variation. Perhaps the reason we had no problem with the fermion is related to the fact that the ten dimensional Yang-Mills fermion variation contains terms of order $F^3$ whereas the gauge-field variation contains terms of order $F^2$. The result being that the dualised fermion variation contains terms with the \textit{same} structure that derive from \textit{different} ten-dimensional terms. This allows for the coefficients to be related in such a way that unwanted terms are eliminated. This is not true for the scalars. Furthermore, by observing how the terms in \eqref{bose} break-up into SO(7) objects
\begin{align}
\be \g^I \psi \partial_\mu X^J \partial^\mu X^J &\ra \be \g^i \psi \partial_\mu X^8 \partial^\mu X^8 + \be \g^i \psi \partial_\mu X^j \partial^\mu X^j \nn \\
\be \g^{\mu \nu}\g^{IJK} \psi \partial_\mu X^J \partial_\nu X^K &\ra  2\be \g^{\mu \nu} \g^{ij} \psi \partial_\mu X^j \partial_\nu X^8 + \be \g^{\mu \nu} \g^{ijk} \psi \partial_\mu X^j \partial_\nu X^k \nn \\
\be \g^J \psi \partial_\mu X^I \partial^\mu X^J &\ra \be \g^8 \psi \partial_\mu X^i \partial^\mu X^8 + \be \g^j \psi \partial_\mu X^i \partial^\mu X^j \nn \\
\be \g^J \g^{\mu \nu} \psi \partial_\mu X^I \partial_\nu X^J &\ra \be \g^8 \g^{\mu \nu} \psi \partial_\mu X^i \partial_\nu X^8 + \be \g^j \g^{\mu \nu} \psi \partial_\mu X^i \partial_\nu X^j \nn
\end{align}
we see that there are terms appearing in \eqref{failure1} which do not appear in \eqref{bose}. Therefore, until we know how to modify the abelian duality transformation such that the scalar fields combine into SO(8) objects, we will have to follow a different path to determine $\delta X^I$. In the next section we will use our knowledge of the higher order abelian Lagrangian to determine both $\delta \psi$ and $\delta X^I$ by requiring invariance of the action. 
\subsection{Invariance of Higher Order Lagrangian}
The $l_p^3$ corrected abelian M2-brane Lagrangian takes the form \cite{Alishahiha:2008rs, Ezhuthachan:2009sr}
\begin{align}
S_{BLG} = \int d^3 x &- \frac{1}{2} \partial_\mu X^I \partial^\mu X^I + \frac{i}{2} \bp \g^\mu \partial_\mu \psi \nn \\
&+ \frac{1}{4}l_p^3 (\partial^\mu X^I \partial_\mu X^J \partial^\nu X^J \partial_\nu X^I - \frac{1}{2} \partial^\mu X^I \partial_\mu X^I \partial^\nu X^J \partial_\nu X^J) \nn \\
&+ \frac{i}{4}l_p^3 ( \bp \g^\mu \g^{IJ} \partial_\nu \psi \partial_\mu X^I \partial^\nu X^J - \bp \g^\mu \partial_\nu \psi \partial_\mu X^I \partial^\nu X^I) \nn \\
&-\frac{1}{16}  l_p^3 \bp \g^\mu \partial^\nu \psi \bp \g_\nu \partial_\mu \psi
\end{align}
The supersymmetry transformations at lowest order are
\begin{align}
\delta X^I &= i \be \g^I \psi. \nn \\
\delta \psi &= \partial_\mu X^I \g^\mu \g^I \e.
\end{align}
At higher order we will consider the transformations \eqref{fermi} and \eqref{bose} (neglecting bi-linear and tri-linear fermion terms). To recap, for the fermion we have
\begin{align}
\delta \psi = &+ a_1 \g_\mu \g^I \partial_\nu X^J \partial^\nu X^J \partial^\mu X^I \e + a_2 \g_\mu \g^I \partial^\mu X^J \partial_\nu X^J \partial^\nu X^I \e \nn \\
&+ a_3\g^{\mu \nu \rho} \g^{IJK} \partial_\mu X^I \partial_\nu X^J \partial_\rho X^K \e 
\end{align}
and for the scalar 
\begin{align}
\delta X^I =  &+ b_1 \be \g^I \psi \partial_\mu X^J \partial^\mu X^J + b_2 \be \g^J \psi \partial_\mu X^I \partial^\mu X^J \nn \\
&+ b_3 \be \g^J \g^{\mu \nu} \psi \partial_\mu X^I \partial_\nu X^J + b_4 \be \g^{\mu \nu} \g^{IJK} \psi \partial_\mu X^J \partial_\nu X^K.
\end{align}
In the variation of the action there will be terms coming from the higher order supersymmetry variation of the lower order Lagrangian and there will be terms coming from the lower order supersymmetry variation of the higher order Lagrangian. These terms should cancel against each other up to a surface term. Demanding invariance of the action will put constraints on the coefficients. Not only will we determine $\delta X^I$ but also $\delta \psi$ allowing for comparison with the result derived in the previous section using abelian duality. Let us begin by considering the higher order supersymmetry variation of the lower order Lagrangian. This results in
\begin{align}
- \partial_\mu X^I \partial^\mu (\delta X^I) = &-i b_1 \be \g^I \partial_\mu \psi \partial^\mu X^I \partial^\nu X^J \partial_\nu X^J  - i b_1 \be \g^I \psi \partial_\mu (\partial_\nu X^J \partial^\nu X^J) \partial^\mu X^I \nn \\
&- i b_2 \be \g^I \partial_\mu \psi \partial^\mu X^J \partial^\nu X^J \partial_\nu X^I - i b_2 \be \g^I \psi \partial_\mu (\partial^\nu X^J \partial_\nu X^I) \partial^\mu X^J \nn \\
&- i b_3 \be \g^I \g^{\mu \nu} \partial_\rho \psi \partial_\mu X^J \partial_\nu X^I \partial^\rho X^J - i b_3 \be \g^I \g^{\mu \nu} \psi \partial_\rho (\partial_\mu X^J \partial_\nu X^I) \partial^\rho X^J \nn \\
&- i b_4 \be \g^{\rho \nu} \g^{IJK} \partial_\mu \psi \partial^\mu X^I \partial_\rho X^J \partial_\nu X^K  - i b_4 \be \g^{\rho \nu } \g^{IJK} \psi \partial_\mu (\partial_\rho X^J \partial_\nu X^K) \partial^\mu X^I 
\end{align}
\begin{align}
\delta (\frac{i}{2} \bp \g^\mu \partial_\mu \psi) = &+ \frac{i}{2} a_1 \be \g^I \partial_\mu \psi \partial^\mu  X^I \partial_\nu X^J \partial^\nu X^J - \frac{i}{2} a_1 \be \g^I \psi \partial_\mu (\partial_\nu X^J \partial^\nu X^J \partial^\mu X^I)  \nn \\
&+ \frac{i}{2} a_2 \be \g^I \partial_\mu \psi \partial^\mu X^J \partial_\nu X^J \partial^\nu X^I - \frac{i}{2} a_2 \be \g^I \psi \partial_\mu (\partial^\mu X^J \partial_\nu X^J \partial^\nu X^I )  \nn \\
&- \frac{i}{2} a_1 \be  \g^I \g^{\mu \nu} \partial_\mu \psi \partial_\nu X^I \partial^\rho X^J \partial_\rho X^J + \frac{i}{2} a_1 \be  \g^I \g^{\mu \nu} \psi \partial_\mu (\partial_\rho X^J \partial^\rho X^J \partial_\nu X^I)  \nn \\
&- \frac{i}{2} a_2 \be  \g^I \g^{\mu \nu} \partial_\mu \psi \partial_\nu X^J \partial^\rho X^J \partial_\rho X^I + \frac{i}{2} a_2 \be \g^I \g^{\mu \nu} \psi \partial_\mu (\partial_\nu X^J \partial^\rho X^J \partial_\rho X^I)  \nn \\
&- \frac{3i}{2} a_3 \be \g^{\rho \nu } \g^{IJK} \partial^\mu \psi \partial_\mu X^I \partial_\nu X^J \partial_\rho X^K  + \frac{3i}{2} a_3 \be \g^{\rho \nu } \g^{IJK} \psi \partial_\mu (\partial^\mu X^I \partial_\nu X^J \partial_\rho X^K) 
\end{align}
Let us now look at the lower order supersymmetry variation of the higher-order Lagrangian terms. We have
\begin{align}
\delta {\mathcal{L}}_{higher} = &+i \be \g^I \partial_\mu \psi \partial^\mu X^J \partial_\nu X^J \partial^\nu X^I - \frac{i}{2} \be \g^I \partial^\mu \psi \partial_\mu X^I \partial^\nu X^J \partial_\nu X^J \nn \\
&+ \frac{i}{4}  \be \g^I \partial_\mu \psi \partial^\mu X^I \partial_\nu X^J \partial^\nu X^J - \frac{i}{2} (2 \pi )^2 \be \g^I \partial_\mu \psi \partial^\mu X^J \partial_\nu X^J \partial^\nu X^I \nn \\
&- \frac{i}{4}  \be \g^{\mu \nu} \g^I \psi \partial_\mu X^J \partial^\rho X^I (\partial_\nu \partial_\rho X^J ) + \frac{i}{4}  \be \g^{\mu \nu} \g^I \psi  \partial_\mu X^I \partial^\rho X^J (\partial_\nu \partial_\rho X^J) \nn \\
&- \frac{i}{4}  \be \g^{\mu \nu} \g^I \psi \partial_\mu X^J \partial^\rho X^J (\partial_\nu \partial_\rho X^I) + \frac{i}{4} \be \g^I \psi \partial^\mu X^J \partial^\rho X^J (\partial_\mu \partial_\rho X^I ) \nn \\
&- \frac{i}{4} \be \g^{\rho \nu} \g^{IJK} \psi \partial_\mu X^I \partial_\nu X^J (\partial^\mu \partial_\rho X^K) + \frac{i}{4} \be \g^{\rho \nu} \g^{IJK} \partial_\mu \psi \partial^\mu X^I \partial^\rho X^J \partial_\nu X^K \nn \\
&+ \psi^3  \textrm{terms}
\end{align}
How will these terms cancel against each other? Firstly we observe that there are three `types' of term appearing in the above, depending on the gamma matrix structure. It is clear that terms involving the same gamma matrix structure should cancel against each other (up to total derivatives). We begin by focusing on $\g^{\nu \rho} \g^{IJK}$ terms.
Collecting these terms together we can write them as
\begin{align}
&(\frac{3i}{2} a_3 - i b_4 + \frac{i}{4} ) \be \g^{\rho \nu} \g^{IJK} \partial_\mu \psi \partial^\mu X^I \partial_\rho X^J \partial_\nu X^K - \frac{3i}{2} a_3 \be \g^{\rho \nu} \g^{IJK} \psi \partial_\mu (\partial^\mu X^I \partial_\nu X^K \partial_\rho X^J) \nn \\
&+ 2i b_4 \be \g^{\rho \nu} \g^{IJK} \partial_\mu X^I \partial_\nu X^J (\partial^\mu \partial_\rho X^K) \psi - \frac{i}{4}  \be \g^{\rho \nu} \g^{IJK} \psi \partial_\mu X^I \partial_\nu X^J (\partial^\mu \partial_\rho X^K) 
\end{align}
We notice that the first line can be expressed as a total derivative and the second line vanishes provided that
\begin{align}
b_4 &= \frac{1}{8} , \qquad a_3 = -\frac{1}{24} 
\end{align}
Now let us look at the $\g^I$ terms. We can write these as
\begin{align}
&+i(\frac{1}{2} a_1 - b_1 - \frac{1}{4}) \be \g^I \partial^\mu \psi \partial_\mu X^I \partial^\nu X^J \partial_\nu X^J - \frac{i}{2} a_1 \be \g^I \psi \partial_\mu (\partial_\nu X^J \partial^\nu X^J \partial^\mu X^I) \nn \\
&+i (\frac{1}{2} a_2 - b_2 + \frac{1}{2}) \be \g^I \partial_\mu \psi \partial^\mu X^J \partial_\nu X^J \partial^\nu X^I - \frac{i}{2} a_2 \be \g^I \psi \partial_\mu (\partial^\mu X^J \partial_\nu X^J \partial^\nu X^I ) \nn \\
&+ \frac{i}{4} \be \g^I \psi \partial^\mu X^J \partial^\nu X^J (\partial_\mu \partial_\nu X^I) - i b_2 \be \g^I \psi \partial^\nu X^J \partial^\mu X^J (\partial_\mu \partial_\nu X^I) \nn \\
&-i b_2 \be \g^I \psi (\partial_\mu \partial_\nu X^J) \partial^\nu X^I \partial^\mu X^J - 2 i b_1 \be \g^I (\partial_\mu \partial_\nu X^J) \partial^\nu X^J \partial^\mu X^I \label{bzo}
\end{align}
The first two lines can be expressed as total derivatives provided that
\begin{align}
a_1 &= b_1 + \frac{1}{4}  \nn \\
a_2 &= b_2 - \frac{1}{2} 
\end{align}
We see that the last two lines in \eqref{bzo} are equal to zero provided
\begin{align}
b_1 = -\frac{1}{8}, \qquad b_2 = \frac{1}{4} 
\end{align}
Putting this information together we conclude 
\begin{equation}
a_2 = - 2 a_1. \label{boya}
\end{equation}
This agrees with the result derived from the duality transformation method. Now we just need to check that the remaining terms cancel against each other. Focusing on the $\g^J \g^{\rho \mu}$ terms we see they can be written as
\begin{align}
&- b_3 l_p^3 \be \g^J \g^{\rho \mu} \partial_\nu \psi \partial_\rho X^I \partial_\mu X^J \partial^\nu X^I + \frac{i}{2} a_1 l_p^3 \be \g^J \g^{\rho \mu} \partial_\mu \psi \partial_\nu X^I \partial^\nu X^I \partial_\rho X^J \nn \\
&+ \frac{i}{2} l_p^3 a_2 \be \g^J \g^{\rho \mu} \partial_\mu \psi \partial_\rho X^I \partial^\nu X^I \partial_\nu X^J + (\frac{i}{4} + b_3 + \frac{i}{2} a_2 ) l_p^3 \be \g^J \g^{\rho \mu} \psi (\partial_\nu \partial_\rho X^J) \partial_\mu X^I \partial^\nu X^I  \nn \\
&+ (i a_1 - \frac{i}{4} - b_3 ) l_p^3 \be \g^J \g^{\rho \mu} \psi (\partial_\nu \partial_\rho X^I ) \partial^\nu X^I \partial_\mu X^J + (\frac{i}{4} + \frac{i}{2} a_2 ) l_p^3 \be \g^J \g^{\rho \mu} \psi (\partial_\nu \partial_\rho X^I) \partial_\mu X^I \partial^\nu X^J 
\end{align}
If we set $b_3 = 0$ then it is possible to write the remaining terms as total derivatives provided that
\begin{align}
a_1 = \frac{1}{8}, \qquad a_2 = - \frac{1}{4}
\end{align}
which is consistent with \eqref{boya}. We are now in a position to write down expressions for the $\mathcal{O} (l_p^3)$ corrections to the abelian supersymmetry transformations (excluding bi-linear and tri-linear fermion terms)
\begin{align}
\delta X^I &= i \be \g^I \psi - \frac{i}{8} l_p^3 \be \g^I \psi \partial_\mu X^J \partial^\mu X^J + \frac{i}{4} l_p^3 \be \g^J \psi \partial_\mu X^I \partial^\mu X^J + \frac{i}{8} l_p^3 \be \g^{\mu \nu} \g^{IJK} \psi \partial_\mu X^J \partial_\nu X^K. \nn \\
\delta \psi &= \partial_\mu X^I \g^\mu \g^I \e + \frac{1}{8}l_p^3 \g^\mu \g^I \partial_\mu X^I \partial^\nu X^J \partial_\nu X^J\e - \frac{1}{4}l_p^3  \g^\mu \g^I \partial_\mu X^J \partial^\nu X^J \partial_\nu X^I \e \nn \\
&\quad - \frac{1}{24} l_p^3 \g^{\mu \nu \rho} \g^{IJK} \partial_\mu X^I \partial_\nu X^J \partial_\rho X^K \e.
\end{align}
Looking at the fermion variation we see that it is possible to fix the undetermined overall coefficient in \eqref{magic2} as $\lambda_1 = \frac{1}{32}$. In the next section we will consider extending this analysis to the non-abelian Bagger-Lambert M2-brane theory.
\section{Non-Abelian Extension}
In this section we begin an investigation into the non-abelian supersymmetry transformation of the Bagger-Lambert theory at $\mathcal{O} (l_p^3)$. We will see that using the non-abelian dNS duality transformation outlined at the beginning of the paper it is possible to uniquely determine the higher order fermion variation. We begin by using dimensional analysis to determine the types of terms that can appear in the M2-brane supersymmetry transformations.
\subsection{Dimensional Analysis}
We can write the most general variation of the fermion field as
\begin{equation}
\delta \psi_a = D_\mu X^I_a \g^\mu \g^I \e + \frac{1}{6} X^{IJK} \g^{IJK} \e + l^3_p D^m X^n \psi^{2l} \e.
\end{equation}
Dimensional analysis then tells us that
\begin{equation}
2m + n + 4l = 9
\end{equation}
This gives rise to potentially nine types of term. However we can restrict our attention by making use of our knowledge of the D2-brane supersymmetry transformations. In other words we will only consider terms that match the D2-brane corrections upon application of the novel Higgs mechanism. This leaves us with
\begin{enumerate}
\item $X^9$ .
\item $(DX)X^6$.
\item $(DX) (DX) X^3$.
\item  $(DX)(DX)(DX)$.
\end{enumerate}
working out all independent index contractions one finds an expression of the form
\begin{align}
\delta \psi = l_p^3 [ &a_1 \g_\mu \g^I  D_\nu X^J D^\nu X^J D^\mu X^I + a_2 \g_\mu \g^I  D^\mu X^J D_\nu X^J D^\nu X^I + a_3 \e^{\mu \nu \rho} \g^{IJK}  D_\mu X^I D_\nu X^J D_\rho X^K \nn \\
&+ a_4 \g^{\mu \nu} \g^I  D_\mu X^J D_\nu X^K X^{JKI}  + a_5 \g^{IJK}   D_\mu X^L D^\mu X^J X^{ILK} + a_6 \g^{IJK}  D_\mu X^L D^\mu X^L X^{IJK} \nn \\
&+ a_7 \g^{\mu \nu} \g^{IJKLM}  D_\mu X^I D_\nu X^J X^{KLM} + a_8 \g_\mu \g^J  D^\mu X^K X^{KLM} X^{LJM} \nn \\
&+ a_9 \g_\mu \g^{IJKLM}  D^\mu X^M X^{IJN} X^{KLN} + a_{10} \g_\mu \g^J D_\mu X^J X^{KLM} X^{KLM}  \nn \\
&+ a_{11} \g_\mu \g^{IJKLM} D^\mu X^N X^{IJM} X^{KLN} + a_{12} \g^{IJKLMNP}  X^{IJQ} X^{KLQ} X^{MNP}  \nn \\
&+ a_{13}\g^{IJM}  X^{IKN} X^{KLN} X^{LJM} +  a_{14} \g^{IJM} X^{KLN} X^{KLN} X^{IJM} ]\e. \label{hf}
\end{align}
Similarly we can write the most general scalar field variation as
\begin{equation}
\delta X^{I}_a = i\be \g^I \psi_a + l^3_p \be D^m X^n \psi^{2l+1}
\end{equation}
with
\begin{align}
2m + n + 4l = 6.
\end{align}
This leads to the following possible terms
\begin{enumerate}
\item $\psi X^6$.
\item $\psi (DX)X^3$.
\item $\psi (DX)(DX)$.
\end{enumerate}
After a little thought about possible index contractions one arrives at the following expression
\begin{align}
\delta X^I = il_p^3 [&b_1 \be \g^I \psi D_\mu X^J D^\mu X^J + b_2 \be \g^J \psi D_\mu X^I D^\mu X^J  + b_3 \be \g^J \g^{\mu \nu} \psi D_\mu X^I D_\nu X^J   \nn \\
&+ b_4 \be \g^{\mu \nu}\g^{IJK} \psi D_\mu X^J D_\nu X^K + b_5 \be \g_\mu \g^{JKL} \psi D^\mu X^I X^{JKL} + b_6 \be \g_\mu \g^{IJK} \psi D^\mu X^L X^{JKL} \nn \\
&+ b_7 \be \g_\mu \g^J \psi D_\mu X^K X^{IJK} + b_8 \be \g^\mu \g^{IJKLM} \psi D_\mu X^J X^{KLM}  + b_9 \be \g_\mu \g^{JKL} \psi D^\mu X^K X^{IJL} \nn \\ 
&+ b_{10} \be \g^J \psi X^{JKL} X^{IKL}  + b_{11} \be \g^{JKL} \psi X^{KLN} X^{NIJ}  + b_{12} \be \g^{IJKLM} \psi X^{JKN} X^{LMN}]. \label{hj}
\end{align}
Now that we know the types of terms that will appear in the supersymmetry transformations we will use the non-abelian dNS duality transformation outlined at the beginning of the chapter to try and determine their exact form. Our starting point is the non-abelian D2-brane supersymmetry transformations.
\subsection{D2-brane supersymmetry transformations}
We begin by deriving the non-abelian D2-brane $\alpha'^2$ supersymmetry transformations. Our starting point will be the $\alpha'^2$ ten-dimensional $U(N)$ super Yang-Mills transformations \eqref{ymtran}. Because we are now considering the full non-abelian theory we will have to keep all terms in the dimensional reduction, including commutator terms. Upon dimensional reduction to (2+1) dimensions one finds the following expressions
\begin{equation}
\delta X^i = \sum_{j=1}^6  \delta X^i_{(j)} \label{uno}
\end{equation}
with
\begin{align}
\delta X^i_{(1)} &= \frac{1}{g^2_{YM}}\alpha_1 \be \g^i \psi F_{\mu \nu} F^{\mu \nu} \nn \\
\delta X^i_{(2)} &= -\frac{1}{g_{YM}} (\alpha_2 \be \g_\mu \psi D_\rho X^i F^{\rho \mu} - \alpha_3 \e_{\mu \rho \sigma}\be  \psi D^\mu X^i F^{\rho \sigma} - 4 \alpha_4 \e_{\mu \nu \rho} \be \g^{ij}  \psi F^{\mu \nu} D^\rho X^j )\nn \\
\delta X^i_{(3)} &= - \alpha_3 \be \g^{\rho \sigma} \g^j X^{ij} F_{\rho \sigma} - 2 \alpha_4 \be \g^{ijk} \g_{\mu \nu} \psi X^{jk} F^{\mu \nu} \nn \\
\delta X^i_{(4)} &= 2 \alpha_1 \be \g^i \psi D_\mu X^j D^\mu X^j - \alpha_2 \be \g^j \psi D_\rho X^i D^\rho X^j - \alpha_3 \be \g^{\mu \rho} \g^j \psi D_\mu X^i D_\rho X^j \nn \\
&- \alpha_3 \be \g^{\mu \sigma} \g^j \psi D_\mu X^i D_\sigma X^j - 4 \alpha_4 \be \g^{ijk} \g_{\mu \nu} \psi D_\mu X^j D_\nu X^k \nn \\
\delta X^i_{(5)} &= g_{YM} (\alpha_2 \be \g_\mu \psi X^{ij} D_\mu X^j + \alpha_3 \be \g^\mu \g^{jk} \psi D_\mu X^i X^{jk} + \alpha_3 \be \g^\rho \g^{jk} \psi  X^{ij} D_\rho X^k \nn \\
&+ \alpha_3 \be \g^\sigma \g^{jk} \psi X^{ij} D_\sigma X^k + 4 \alpha_4 \be \g_\mu \g^{ijkl} \psi D^\mu X^j X^{kl}) \nn \\
\delta X^i_{(6)} &= g^2_{YM} (\alpha_1 \be \g^i \psi X^{jk} X^{jk} + \alpha_2 \be \g^k \psi X^{ij} X^{jk} + \alpha_3 \be \g^{jkl} \psi X^{ij} X^{kl} + \alpha_4 \be \g^{ijklm} \psi X^{jk} X^{lm} )\nn
\end{align}
\\
\\
\begin{equation}
\delta A_\mu = \sum_{i=1}^6  \delta A_{\mu (i)}  \label{duz}
\end{equation}
with
\begin{align}
\delta A_{\mu (1)} &=  \frac{1}{g_{YM}}(\alpha_1 \be \g_\mu \psi F_{\rho \sigma} F^{\rho \sigma} + \alpha_2 \be \g_\rho \psi F_{\mu \nu} F^{\nu \rho} - \alpha_3 \e^{\nu \rho \sigma} \be \psi F_{\mu \nu} F_{\rho \sigma}) \nn \\
\delta A_{\mu (2)} &= \alpha_2 \be \g^j \psi F_{\mu \nu} D^\nu X^j + \alpha_3 \be \g^{\nu \sigma} \g^j \psi F_{\mu \nu} D_\sigma X^j + \alpha_3 \be \g^{\rho \sigma} \g^j \psi D_\mu X^j F_{\rho \sigma} + \alpha_3 \be \g^{\nu \rho} \g^k \psi F_{\mu \nu} D_\rho X^k \nn \\
\delta A_{\mu (3)} &= - g_{YM} (\alpha_3 \be \g^\nu \g^{jk} \psi F_{\mu \nu} X^{jk}
 - \alpha_4 \e_{\mu \nu \rho} \be \g^{ij} \psi F^{\nu \rho} X^{ij} - \alpha_4 \e_{\mu \nu \rho} \be  \g^{ij} \psi X^{ij} F^{\nu \rho} )\nn \\
 \delta A_{\mu (4)} &= g_{YM} (2 \alpha_1 \be \g_\mu \psi D_\nu X^i D^\nu X^i - \alpha_2 \be \g_\nu \psi D_\mu X^i D^\nu X^i - \alpha_3 \be \g^{ij} \g^\sigma \psi D_\mu X^i D_\sigma X^j  \nn \\
 &- \alpha_3 \be \g^{ij} \g^\rho \psi D_\mu X^i D_\rho X^j + 4 \alpha_4 \e_{\mu \nu \rho} \be \g^{ij} \psi D^\nu X^i D^\rho X^j ) \nn \\ 
\delta A_{\mu (5)} &= - g^2_{YM} (\alpha_2 \be \g^j \psi D_\mu X^i X^{ij} + \alpha_3 \be \g^{ijk} \psi D_\mu X^i X^{jk} \nn \\
&+ 2 \alpha_4 \be \g_{\mu \nu} \g^{jkl} \psi D^\nu X^j X^{kl} + 2 \alpha_4 \be \g_{\mu \nu} \g^{jkl} \psi X^{jk} D^\nu X^l ) \nn \\
\delta A_{\mu (6)} &= g^3_{YM} (\alpha_1 \be \g_\mu \psi X^{ij} X^{ij} + \alpha_4 \be \g_\mu \g^{ijkl} \psi X^{ij} X^{kl}) \nn
\end{align}

\begin{equation}
\delta \psi = \sum_{i=1}^{10}  \delta \psi_i \e \label{trez}
\end{equation}
with
\begin{align}
\delta \psi_1 &= \frac{1}{g^3_{YM}}(\lambda_1 \g^{\mu \nu} F_{\rho \sigma} F^{\rho \sigma} F_{\mu \nu} + \lambda_2 \g^{\mu \nu} F_{\mu \rho} F^{\rho \sigma} F_{\sigma \nu})  \nn \\
\delta \psi_2 &= \frac{1}{g^2_{YM}} (2 \lambda_1 \g^\mu \g^j F_{\rho \sigma} F^{\rho \sigma} D_\mu X_j + \lambda_2 \g^\mu \g^j F_{\mu \nu} F^{\nu \rho} D_\rho X_j +  \lambda_2 \g^\nu \g^j D_\rho X_j F^{\rho \sigma} F_{\sigma \nu} ) \nn \\
\delta \psi_3 &= - \frac{1}{g_{YM}}(\lambda_1 \g^{ij} F_{\rho \sigma} F^{\rho \sigma} X^{ij}) \nn \\
\delta \psi_4 &= \frac{1}{g_{YM}} ( \g^{\mu \nu}(2 \lambda_1  D_\rho X^j D^\rho X^j F_{\mu \nu} - \lambda_2  D_\mu X^j D^\sigma X^j F_{\sigma \nu}- \lambda_2 F_{\mu \rho} D^\rho X^k D_\nu X^k) \nn \\
&\quad - \lambda_2 \g^{ij} D_\rho X^i F^{\rho \sigma} D_\sigma X_j )\nn \\
\delta \psi_5 &= -\lambda_2 \g^\mu \g^j F_{\mu \nu} D^\nu X^k X^{kj} - \lambda_2 \g^\nu \g^i X^{ij} D^\rho X^j F_{\rho \nu} - 12 \lambda_3 \e^{\mu \nu \rho} \g^{ijk} F_{\mu \nu} D_\rho X^i X^{jk} \nn \\
\delta \psi_6 &= g_{YM} (\lambda_1 \g^{\mu \nu} X^{ij} X^{ij} F_{\mu \nu} + 3 \lambda_3 \g^{\mu \nu} \g^{ijkl} F_{\mu \nu} X^{ij} X^{kl} ) \nn \\
\delta \psi_7 &= 4 \lambda_1 \g^\mu \g^j D_\nu X^k D^\nu X^k D_\mu X^j - \lambda_2 \g^\mu \g^j D_\mu X^k D^\rho X^k D_\rho X^j \nn \\ &- \lambda_2 \g^\nu \g^i D_\rho X^i D^\rho X^j D_\nu X^j - 8 \lambda_3 \e^{\mu \nu \rho} \g^{ijk} D_\mu X^i D_\nu X^j D_\rho X^k \nn \\
\delta \psi_8 &= g_{YM} (-2 \lambda_1 \g^{ij} D_\mu X^k D^\mu X^k X^{ij} + \lambda_2 \g^{\mu \nu} D_\mu X^j X^{jk} D_\nu X^k + \lambda_2 \g^{ij} D_\rho X^i D^\rho X^k X^{kj} \nn \\
&+ \lambda_2 \g^{ij} X_{ik} D^\rho X^k D_\rho X_j + 12 \lambda_3 \g^{\mu \nu} \g^{ijkl} D_\mu X^i D^\nu X^j X^{kl} )\nn \\
\delta \psi_9 &= g^2_{YM} \g^\mu (2 \lambda_1  \g^j X^{kl} X^{kl} D_\mu X^j + \lambda_2 \g^i X^{ij} X^{jk} D_\mu X^k + \lambda_2  \g^j D_\mu X^k X^{kl} X^{lj} \nn \\
&\quad +  6 \lambda_3  \g^{ijklm} D_\mu X^i X^{jk} X^{lm}) \nn \\
\delta \psi_{10} &= - g^3_{YM} (\lambda_1 \g^{ij} X^{kl} X^{kl} X^{ij} + \lambda_2 \g^{ij} X^{ik} X^{kl} X^{lj} + \lambda_3 \g^{ijklmn} X^{ij} X^{kl} X^{mn}). \nn
\end{align}
Now that we have the non-abelian D2-brane supersymmetry transformations we can attempt to dualise the Yang-Mills gauge field to a scalar. Our method will follow the presentation of \cite{Alishahiha:2008rs} where the $l_p^3$ corrections to the Lorentzian BLG theory were derived using the dNS duality prescription. Let us briefly review the procedure for implementing dNS duality in the higher order Lagrangian.
\subsection{Higher order dNS duality}
In \cite{Alishahiha:2008rs},  higher order corrections to the Lorentzian BLG theory were derived by making use of the dNS duality transformation. As outlined in Section 2, implementing the dNS duality involves rewriting the D2-brane Lagrangian in terms of the new fields $B_\mu$ and $X^8$. To see how this works at higher order we will derive the $\mathcal{O} (l_p^3)$ bosonic terms of the BLG theory. Our starting point will be the $(\alpha')^2$ corrections to the non-abelian D2-brane theory. These terms derive from the $F^4$ corrections of ten dimensional super Yang-Mills theory \cite{Tseytlin:1997csa, Cederwall:2001td, Bergshoeff:2001dc}
\begin{align}
\mathcal{L} &= -\frac{1}{4} F^2 + \frac{1}{8} \Str (F^4 - \frac{1}{4} (F^2)^2) \nn \\
&=-\frac{1}{4}F_{MN}F^{MN} + \frac{1}{12} \Tr [F_{MN} F_{RS} F^{MR} F^{NS} + \frac{1}{2} F_{MN} F^{NR} F_{RS} F^{SM} \nn \\
&\quad \quad - \frac{1}{4} F_{MN} F^{MN} F_{RS} F^{RS} - \frac{1}{8} F_{MN} F_{RS} F^{MN} F^{RS}] .\label{10d}
\end{align} 
The next step is to reduce this expression to (2+1) dimensions. We then re-write the (2+1) dimensional field strength $F_{\mu \nu}$ in terms of the dual field strength ${\tilde{F}}_\mu = \e_{\mu \nu \lambda} F^{\nu \lambda}$. In order to implement the dNS duality we replace the dual field strength ${\tilde{F}}_\mu$ by an independent matrix-valued one-form field $B_\mu$. The resulting Lagrangian looks like
\begin{align}
\mathcal{L} = &\Tr [ {\tilde{F}}_\mu B^\mu - \frac{g^2_{YM}}{2} B_\mu B^\mu + \frac{g^4_{YM}}{4} (B_\mu B^\mu B_\nu B^\nu + \frac{1}{2} B_\mu B_\nu B^\mu B^\nu ) \nn \\
&+ \frac{g^2_{YM}}{12} (2 B^\mu B_\nu D^\nu X^i D_\mu X^i - 2B^\mu B_\mu D_\nu X^i D^\nu X^i + 2 B^\mu B^\nu D_\mu X^i D_\nu X^i \nn \\
&+ B^\mu D^\nu X^i B_\nu D_\mu X^i - B^\mu D^\nu X^i B_\mu D_\nu X^i + B^\mu D_\mu X^i B^\nu D_\nu X^i) \nn \\
&+ \frac{g^4_{YM}}{12} (B^\mu B_\mu X^{ij} X_{ij} + \frac{1}{2} B^\mu X^{ij} B_\mu X^{ij}) \nn \\
&+ \frac{g^2_{YM}}{6} \e_{\mu \nu \lambda} (B^\lambda D^\mu X^i D^\nu X^j + D^\nu X^j B^\lambda D^\mu X^i + D^\mu X^i D^\nu X^j B^\lambda ) X^{ij} ]. \label{poor}
\end{align} 
We see that $\tilde{F}$ only appears in the Chern-Simons term ${\tilde{F}}_\mu B^\mu$. To show that this expression is equivalent to the $(\alpha' )^2$ D2-brane Lagrangian one simply integrates out the field $B_\mu$ order by order using its equation of motion. In order to rewrite the Lagrangian in an SO(8) invariant form we introduce the field $X^8$ and replace $B_\mu$ everywhere it occurs by $- 1/g_{YM} (D_\mu X^8 - g_{YM} B_\mu)$. 
Performing this substitution and collecting the resulting terms into the SO(8) invariant building blocks $\D_\mu X^I$ and $X^{IJK}$ results in the compact expression
\begin{align}
\mathcal{L} = &+\frac{1}{2} \e_{\mu \nu \lambda} B^\mu F^{\nu \lambda} - \frac{1}{2} \D_\mu X^I \D^\mu X^I \nn \\
 &+ \frac{1}{8} l_p^3 \Str [ 2 \D^\mu X^I \D_\mu X^J \D^\nu X^J \D_\nu X^I - \D^\mu X^I \D_\mu X^I \D^\nu X^J \D_\nu X^J  \nn \\
 &- \frac{4}{3} \e^{\mu \nu \lambda} X^{IJK} \D_\mu X^I \D_\nu X^J \D_\lambda X^K \nn \\
 &+ 2 X^{IJK} X^{IJL} \D^\mu X^K \D_\mu X^L - \frac{1}{3} X^{IJK} X^{IJK} \D^\mu X^L \D_\mu X^L \nn \\
 &+ \frac{1}{3} X^{IJM} X^{KLM} X^{IKN} X^{JLN} - \frac{1}{24} X^{IJK} X^{IJK} X^{LMN} X^{LMN} ].
\end{align}
In \cite{Alishahiha:2008rs} it was shown that the same approach can be used to derive the $\mathcal{O} (l_p^3)$ fermion terms. We see that it is possible to implement dNS duality at higher order by applying the following prescription
\begin{enumerate} 
\item Dimensionally reduce 10 dimensional expression to (2+1) dimensions.
\item Write all field strengths in terms of their duals: \quad $F_{\mu \nu} = - \e_{\mu \nu \lambda} {\tilde{F}}^\lambda$.
\item Replace ${\tilde{F}}_\mu$ with the field $B_\mu$.
\item Replace $B_\mu$  with 
$-{g_{YM}} \D_\mu X^8$.
\item Rewrite all expressions in terms of $\D_\mu X^I$ and $X^{IJK}$ building blocks.
\end{enumerate}
In the next section we will test whether this prescription works at the level of supersymmetry transformations. We have already performed the first task on the list. Next we must re-write the D2-brane supersymmetry transformations \eqref{uno}, \eqref{duz} and \eqref{trez} in terms of $\D_\mu X^8$. 

\subsection{dNS transformed supersymmetry}
\begin{align}
\delta X^i = &- \overbrace{2 \alpha_1 \be \g^i \psi \D^\mu X^8 \D_\mu X^8 + \alpha_2 \be \g^{\mu \nu} \psi \D_\mu X^i \D_\nu X^8 -2 \alpha_3 \be \psi \D^\mu X^i \D_\mu X^8}^{\textrm{Two Derivative}} \nn \\
&+2 \alpha_1 \be \g^i \psi \D_\mu X^j \D^\mu X^j - \alpha_2 \be \g^j \psi \D_\mu X^i \D^\mu X^j - 2\alpha_3 \be \g^{\mu \nu} \g^j \psi \D_\mu X^i \D_\nu X^j \nn \\
& - 8 \alpha_4 \be \g^{ij} \psi \D_\mu X^8 \D^\mu X^j - 4 \alpha_4 \be \g^{ijk} \g_{\mu \nu} \psi \D_\mu X^j \D_\nu X^k  \nn \\
&+ \overbrace{ g_{YM}( 2 \alpha_3 \be \g_\mu \g^j X^{ij} \D^\mu X^8 + 4 \alpha_4 \be \g^{ijk} \g^\mu \psi X^{jk} \D_\mu X^8 +\alpha_2 \be \g_\mu \psi X^{ij} \D_\mu X^j }^{\textrm{One Derivative}} \nn \\
&+ \alpha_3 \be \g^\mu \g^{jk} \psi \D_\mu X^i X^{jk} + \alpha_3 \be \g^\rho \g^{jk} \psi  X^{ij} \D_\rho X^k \nn \\
&+ \alpha_3 \be \g^\sigma \g^{jk} \psi X^{ij} \D_\sigma X^k + 4 \alpha_4 \be \g_\mu \g^{ijkl} \psi \D^\mu X^j X^{kl} )\nn \\
&+ \overbrace{ g^2_{YM}( \alpha_1 \be \g^i \psi X^{jk} X^{jk} + \alpha_2 \be \g^k \psi X^{ij} X^{jk} + \alpha_3 \be \g^{jkl} \psi X^{ij} X^{kl} + \alpha_4 \be \g^{ijklm} \psi X^{jk} X^{lm}}^{\textrm{Zero Derivative}}) \label{sc1}
\end{align}

\begin{align}
\delta A_\mu = &+ \overbrace{g_{YM} (- 2 \alpha_1 \be \g_\mu \psi \D_\nu X^8 \D^\nu X^8  + \alpha_2 \be \g_\mu \psi \D^\nu X^8 \D_\nu X^8 - \alpha_2 \be \g_\nu \psi \D^\nu X^8 \D_\mu X^8 }^{\textrm{Two Derivative}} \nn \\
&+ 2 \alpha_3 \e_{\mu \nu \lambda} \be \psi \D^\lambda X^8 \D^\nu X^8   + \alpha_2 \e_{\mu \nu \lambda} \be \g^j \psi \D^\lambda X^8 \D^\nu X^j - 2 \alpha_3 \be \g_\lambda \g^j \psi D_\mu X^j \D^\lambda X^8  \nn \\
&- \alpha_3 \be \g_\mu \g^j \psi \D^\nu X^8 \D_\nu X^j + \alpha_3 \be \g_\lambda \g^j \psi \D^\lambda X^8 \D_\mu X^j - \alpha_3 \be \g_\mu \g^k \psi \D^\rho X^8 \D_\rho X^k  \nn \\
&+ \alpha_3 \be \g_\lambda \g^k \psi \D^\lambda X^8 \D_\mu X^k   + 2 \alpha_1 \be \g_\mu \psi \D_\nu X^i \D^\nu X^i - \alpha_2 \be \g_\nu \psi \D_\mu X^i \D^\nu X^i   \nn \\
 &- \alpha_3 \be \g^{ij} \g^\sigma \psi \D_\mu X^i \D_\sigma X^j   - \alpha_3 \be \g^{ij} \g^\rho \psi \D_\mu X^i \D_\rho X^j +4 \alpha_4 \e_{\mu \nu \rho} \be \g^{ij} \psi \D^\nu X^i \D^\rho X^j )\nn \\
&+ \overbrace{ g^2_{YM} (\alpha_3 \be \g_{\mu \nu} \g^{jk} \psi \D^\nu X^8 X^{jk} - 2 \alpha_4 \be \g^{ij} \psi \D_\mu X^8 X^{ij} - 2 \alpha_4 \be \g_{\mu \nu} \g^{jkl} \psi D^\nu X^j X^{kl}  }^{\textrm{One Derivative}} \nn \\
&- \alpha_2 \be \g^j \psi \D_\mu X^i X^{ij} - \alpha_3 \be \g^{ijk} \psi \D_\mu X^i X^{jk} - 2 \alpha_4 \be \g^{ij} \psi X^{ij} \D_\mu X^8 - 2 \alpha_4 \be \g_{\mu \nu} \g^{jkl} \psi X^{jk} \D^\nu X^l )\nn \\
&+ \overbrace{ g^3_{YM}(\alpha_1 \be \g_\mu \psi X^{ij} X^{ij} + \alpha_4 \be \g_\mu \g^{ijkl} \psi X^{ij} X^{kl} }^{\textrm{Zero Derivative}}) \label{pfm}
\end{align}

\begin{align}
\delta \psi = &+ \overbrace{4 \lambda_1 \g_\mu \e\D^\nu X^8 \D_\nu X^8 \D^\mu X^8 - \lambda_2 \g_\mu \e\D^\nu X^8 \D_\nu X^8 \D^\mu X^8 -\lambda_2 \g_\mu \e \D^\mu X^8 \D_\nu X^8 \D^\nu X^8}^{\textrm{Three Derivative}} \nn \\
&- 4 \lambda_1 \g^\mu \g^j \e \D_\nu X^8 \D^\nu X^8 \D_\mu X^j + \lambda_2 \g^\mu \g^j \e \D^\nu X^8 \D_\nu X^8 \D_\mu X^j - \lambda_2 \g^\mu \g^j \e \D^\nu X^8  \D_\mu X^8 \D_\nu X^j \nn \\
&+ \lambda_2 \g^\mu \g^j \e \D_\mu X^j \D_\nu X^8 \D^\nu X^8 - \lambda_2 \g^\mu \g^j \e \D_\nu X^j \D_\mu X^8 \D^\nu X^8 -4 \lambda_1 \g_\mu \e \D_\nu X^j \D^\nu X^j \D^\mu X^8 \nn \\
& - \lambda_2 \g_\mu \e \D_\nu X^j \D^\mu X^j \D^\nu X^8 + \lambda_2 \g_\mu \e \D_\nu X^j \D^\nu X^j \D^\mu X^8 - \lambda_2 \g_\mu \e \D^\nu X^8 \D^\mu X^k \D_\nu X^k\nn \\
& + \lambda_2 \g_\mu \e \D^\mu X^8 \D^\nu X^k \D_\nu X^k - \lambda_2 \e^{\mu \nu \lambda} \g^{ij} \e \D_\mu X^i \D_\lambda X^8 \D_\nu X^j +4 \lambda_1 \g^\mu \g^j \e \D_\nu X^k \D^\nu X^k \D_\mu X^j \nn \\
& - \lambda_2 \g^\mu \g^j \e \D_\mu X^k \D^\rho X^k \D_\rho X^j - \lambda_2 \g^\nu \g^i \e \D_\rho X^i \D^\rho X^j \D_\nu X^j - 8 \lambda_3 \e^{\mu \nu \rho} \g^{ijk} \e \D_\mu X^i \D_\nu X^j \D_\rho X^k\nn \\
 &+\overbrace{g_{YM}( -\lambda_2 \g_{\mu \nu} \g^j \e \D^\nu X^8 \D^\mu X^k X^{kj} + \lambda_2 \g_{\mu \nu} \g^i \e X^{ij} \D^\mu X^j \D^\nu X^8 + 24 \lambda_3 \g^{ijk} \e \D^\mu X^8 \D_\mu X^i X^{jk}}^{\textrm{Two Derivative}} \nn \\
 & + 2 \lambda_1 \g^{ij} \e \D^\mu X^8 \D_\mu X^8 X^{ij} - 2 \lambda_1 \g^{ij} \e \D_\mu X^k \D^\mu X^k X^{ij} + \lambda_2 \g^{\mu \nu} \e \D_\mu X^j X^{jk} \D_\nu X^k  \nn \\
& + \lambda_2 \g^{ij} \e \D_\rho X^i \D^\rho X^k X^{kj} + \lambda_2 \g^{ij}\e X_{ik} \D^\rho X^k \D_\rho X_j + 12 \lambda_3 \g^{\mu \nu} \g^{ijkl} \e \D_\mu X^i D^\nu X^j X^{kl}) \nn \\
& +\overbrace { g^2_{YM}  (-2 \lambda_1 \g_\mu \e X^{ij} X^{ij} \D^\mu X^8 - 6 \lambda_3 \g_\mu \g^{ijkl} \e \D^\mu X^8 X^{ij} X^{kl} + 2 \lambda_1 \g^\mu \g^j \e X^{kl} X^{kl} \D_\mu X^j}^{\textrm{One Derivative}} \nn \\
&+ \lambda_2 \g^\nu \g^i \e X^{ij} X^{jk} \D_\nu X^k + \lambda_2 \g^\mu \g^j \e \D_\mu X^k X^{kl} X^{lj} + 6 \lambda_3 \g^\mu \g^{ijklm} \e \D_\mu X^i X^{jk} X^{lm}) \nn \\
&- \overbrace{ g^3_{YM} (\lambda_1 \g^{ij} \e X^{kl} X^{kl} X^{ij} + \lambda_2 \g^{ij} \e X^{ik} X^{kl} X^{lj} + \lambda_3 \g^{ijklmn} \e X^{ij} X^{kl} X^{mn} )}^{\textrm{Zero Derivative}}.  \label{massive}
\end{align}

\section{SO(8) supersymmetry transformations}
In the previous section we applied the dNS prescription to the non-abelian D2-brane supersymmetry transformations. We would now like to re-write these expressions in SO(8) form. We will see that this is only possible for the fermion supersymmetry transformation. The scalar transformation is plagued by the same problems we encountered in the abelian theory. We will end this section with a discussion of how one might go about determining the scalar supersymmetry transformation.

\subsection{$\delta \psi$}
Earlier in this chapter we were able to determine the abelian supersymmetry transformation of the fermion by using abelian duality in (2+1) dimensions. In the process we were able to fix the coefficients appearing in \eqref{fermi}. Looking at \eqref{hf} we see that the first three terms are exactly the same as the terms appearing in \eqref{fermi} but with partial derivatives replaced by covariant derivatives. As a result we find that the coefficients are related in exactly the same way. Knowledge of the relationship between $\lambda_1$, $\lambda_2$ and $\lambda_3$, namely
\begin{equation}
\lambda_2 = 4 \lambda_1 ; \quad \lambda_2 = - 24 \lambda_3
\end{equation} 
allows us to re-write all the coefficients in \eqref{massive} in terms of $\lambda_1$. Furthermore by looking at the invariance of the higher order abelian Lagrangian we were able to fix $\lambda_1 = \frac{1}{32}$. Making use of this information, as well as the SO(8) relations outlined in the appendix, it is possible to re-write the two-derivative, one-derivative and zero-derivative terms in \eqref{massive} in an SO(8) invariant form. The final answer for the $l_p^3$ correction to the fermion supersymmetry transformation in BLG theory is
\begin{align}
\delta \psi = l_p^3 [ &\frac{1}{8} \g_\mu \g^I  D_\nu X^J D^\nu X^J D^\mu X^I - \frac{1}{4} \g_\mu \g^I  D^\mu X^J D_\nu X^J D^\nu X^I - \frac{1}{24} \e^{\mu \nu \rho} \g^{IJK}  D_\mu X^I D_\nu X^J D_\rho X^K \nn \\
&+ \frac{1}{8} \g^{\mu \nu} \g^I  D_\mu X^J D_\nu X^K X^{JKI}  + \frac{1}{8} \g^{IJK}   D_\mu X^L D^\mu X^J X^{ILK} - \frac{1}{48} \g^{IJK}  D_\mu X^L D^\mu X^L X^{IJK} \nn \\
& + \frac{1}{48} \g^{\mu \nu} \g^{IJKLM}  D_\mu X^I D_\nu X^J X^{KLM} - \frac{1}{8} \g_\mu \g^J  D^\mu X^K X^{KLM} X^{LJM} \nn \\
&+ \frac{1}{32}\g_\mu \g^{IJKLM}  D^\mu X^M X^{IJN} X^{KLN} + \frac{1}{48} \g_\mu \g^J D_\mu X^J X^{KLM} X^{KLM}  \nn \\
&- \frac{1}{48} \g_\mu \g^{IJKLM} D^\mu X^N X^{IJM} X^{KLN} + \frac{1}{16} \g^{IJKLMNP}  X^{IJQ} X^{KLQ} X^{MNP}  \nn \\
&+ \frac{1}{32}\g^{IJM}  X^{IKN} X^{KLN} X^{LJM} +  \frac{1}{144} \g^{IJM} X^{KLN} X^{KLN} X^{IJM} ]\e.
\end{align}
It is pleasing to see that the dNS duality transformation has allowed us to uniquely determine the structure of the fermion variation. It would be nice to extend this analysis to include tri-linear fermion terms. 

\subsection{$\delta X^I$}
Given that the dNS prescription works for the fermion transformation one might hope that it would also work for the scalar transformation. However, as we observed for the abelian scalar transformation, this is not the case. The two-derivative terms appearing in \eqref{sc1} are of the same form as the abelian scalar terms, with covariant derivatives replacing partial derivatives. For this reason, the non-abelian scalar transformation inherits the same problems we encountered before. For the abelian theory we used a different approach to determine the scalar transformation. This involved checking the invariance of the abelian Lagrangian under a proposed set of supersymmetry transformations (determined by dimensional analysis). The same should be possible for the non-abelian theory. The $l_p^3$ corrected BLG Lagrangian was derived in \cite{Alishahiha:2008rs, Ezhuthachan:2009sr}. Checking that this Lagrangian is invariant (up to surface terms) under the transformations \eqref{hf} and \eqref{hj} should fix the coefficients. Not only would this determine the scalar transformation but would also provide an independent test of the fermion variation calculated using the dNS prescription. 

Ultimately one would like to know how to modify the dNS prescription in such a way that it is possible to derive the scalar variation. Toward this end it may prove useful to determine the scalar transformation by an independent method such that a comparison can be made between the known result and the dNS transformed result \eqref{sc1}. One possibility would be to use the higher order fermion variation to determine the higher order supercharge which could then be used to generate the higher order scalar variation. This should be possible since we observe at lowest order in BLG theory that the supersymmetry current takes the simple form $-\be J^{\mu} = \bp^a \g^\mu \delta \psi_a$ which follows from the fact that the R-current and supersymmetry current reside within the same supersymmetry multiplet. Importantly we see that we only require knowledge of the fermion supersymmetry transformation in order to determine the supersymmetry current. The hope is that a similar relation between fermion variation and supersymmetry current would continue to hold at higher order. 

Another complication worth mentioning is related to the gauge field transformation \eqref{pfm}. For the lower order abelian supersymmetry transformations we observed that the eighth component of the scalar variation $\delta X^I$ arises after dualising $\delta F_{\mu\nu}$. More specifically, looking at \eqref{bt} we see that at lowest order $\partial^\lambda \delta X^8 = i \be \g^8 \partial^\lambda \psi$. In this case, since there is only one field and one derivative on the right-hand side, it is possible to simply `pull off' the derivative to determine $\delta X^8$. This is no longer true at higher order and determining $\delta X^8$ becomes a non-trivial task. 
\section{Outlook}
In this paper we began an investigation into the $l_p^3$ corrections to the BLG supersymmetry transformations. For the abelian theory we were able to determine the the fermion supersymmetry transformation by using an abelian duality transformation. For the scalar transformation we had to use a different approach in which invariance of the higher order abelian Lagrangian was used to fix the coefficients of the transformation. For the non-abelian theory we were able to use the dNS duality transformation to uniquely determine the fermion supersymmetry transformation at $\mathcal{O} (l_p^3)$. It would be interesting to establish the reason why the dNS duality fails to work for the scalar supersymmetry transformation. It should be possible to uniquely determine the form of the $\mathcal{O} (l_p^3)$ scalar transformation by checking the invariance of the higher order Lagrangian derived in \cite{Alishahiha:2008rs, Ezhuthachan:2009sr}. This would also provide an independent check on the fermion result derived using the dNS duality approach. It would also prove interesting to extend this analysis to the $\mathcal{N} = 6$ ABJM theory. Finding such an extension is of great interest as these theories have a clear spacetime interpretation in M-theory. One possibility for how to derive the $\mathcal{N}=6$ result would be to make use of the $\mathcal{N}=8$ result and $SO(8)$ triality. This should work in the same way that it works for the lowest order Bagger-Lambert theory. In \cite{Gustavsson:2009pm} it was shown that the BLG Lagrangian fields could be `triality rotated' in such a way that $({\textbf{8}}_V , {\textbf{8}}_S , {\textbf{8}}_C) \rightarrow ({\textbf{8}}_S , {\textbf{8}}_C , {\textbf{8}}_V)$, where ${\textbf{8}}_V , {\textbf{8}}_S , {\textbf{8}}_C$ are the vector, spinor and cospinor representations of SO(8) respectively. After performing this transformation it is possible to break $SO(8) \rightarrow SU(4)\times U(1)$ and decompose the $SO(8)$ spinor and cospinor fields (and gamma matrices) in order to rewrite the original $\mathcal{N} = 8$ expression in terms of ABJM fields and so-called `non-ABJM' fields. In \cite{Gustavsson:2009pm} the non-ABJM terms were shown to vanish as a result of certain algebraic constraints (deriving from the flatness condition of the gauge field strength). It would be interesting to see whether this analysis can be extended to higher order and if so, whether additional algebraic constraints would be necessary to eliminate the higher order `non-ABJM' terms.

\section*{Acknowledgements\markboth{Acknowledgements}{Acknowledgements}} 
I would like to thank the hospitality of the Tata Institute for Fundamental Research in Mumbai where much of this work was completed. In particular I would like to thank Shiroman Prakash and R. Loganayagam for all of their help and encouragement with this project. I would also like to thank Sunil Mukhi, David Berman, Daniel Thompson and Costis Papageorgakis for useful and enlightening discussions. AL is supported by an STFC grant.

\begin{appendix}
\appendix

\section{Higher order $SO(8)$ invariant objects}
In this section we list $SO(8)$ invariant combinations which give rise to terms appearing in the dNS transformed superymmetry transformations of the previous section. Note that we have suppressed the symmetrised trace in all the expressions that follow.
\subsection{$\delta \psi$}
\subsubsection{Zero Derivative}
\begin{align}
\g^{IJM} X^{KLN} X^{KLN} X^{IJM} &\rightarrow 9g^3_{YM}\g^{ij} X^{kl} X^{kl} X^{ij}  \nn \\
\g^{IJM} X^{IKN} X^{KLN} X^{LJM} &\rightarrow g^3_{YM}\g^{ij} (4X^{ik} X^{kl} X^{lj} - X^{kl} X^{kl} X^{ij}) \nn \\
\g^{IJKLMNP} X^{IJQ} X^{KLQ} X^{MNP} &\rightarrow 3 g^3_{YM} \g^{ijklmn} X^{ij} X^{kl} X^{mn}  \nn 
\end{align}
\subsubsection{One Derivative}
\begin{align}
\g_\mu \g^{J} X^{KLM} X^{KLM} D^\mu X^J &\rightarrow 3 g^2_{YM}(\g_\mu \g^8 X^{kl} X^{kl} D^\mu X^8 +  \g_\mu \g^j X^{kl} X^{kl} D^\mu X^j )\nn \\
\g_\mu \g^J D_\mu X^K X^{KLM} X^{LJM} &\rightarrow g^2_{YM}(2 \g_\mu \g^j D^\mu X^k X^{kl} X^{lj} - \g_\mu \g^8 D^\mu X^8 X^{lm} X^{lm}) \nn \\
\g^\mu \g^J X^{JKM} X^{KLM} D_\mu X^L &\rightarrow g^2_{YM}( 2 \g^\mu \g^j X^{jk} X^{kl} D_\mu X^l - \g^\mu \g^8 X^{km} X^{km} D_\mu X^8) \nn \\
\g_\mu \g^{IJKLM} &\rightarrow g^2_{YM} (\g_\mu \g^{ijklm} D^\mu X^m X^{ij} X^{kl} + \g_\mu \g^{ijkl} D^\mu X^8 X^{ij} X^{kl}) \nn\\
\g_\mu \g^{IJKLM} D^\mu X^N X^{IJM} X^{KLN} &\rightarrow 3 g^2_{YM} \g_\mu \g^{ijkl} D^\mu X^8 X^{ij} X^{kl} \nn \\
\g_\mu \g^{ijklm} D^\mu X^m X^{ij} X^{kl} &\leftarrow g^2_{YM}\g_\mu \g^{IJKLM} (D^\mu X^M X^{IJN} X^{KLN} - \frac{1}{3} D^\mu X^N X^{IJM} X^{KLN}) \nn
\end{align}
\subsubsection{Two Derivative}
\begin{align}
\g^{IJK} D_\mu X^L D^\mu X^L X^{IJK} &\rightarrow 3g_{YM}( \g^{ij} D_\mu X^k D^\mu X^k X^{ij} +  \g^{ij} D_\mu X^8 D^\mu X^8 X^{ij} )\nn \\
\g^{\mu \nu} \g^{M} D_\mu X^K X^{KLM} D_\nu X^L &\rightarrow g_{YM}\g^{\mu \nu}( \g^i D_\mu X^k X^{ik} D_\nu X^8 \nn  +  \g^i D_\mu X^8 X^{li} D_\nu X^l \nn \\
 &+ \g^8 D_\mu X^k X^{kl} D_\nu X^l) \nn \\
\g^{IJK} D_\mu X^I D^\mu X^L X^{LJK} &\rightarrow g_{YM} (\g^{ijk} D_\mu X^i D^\mu X^8 X^{jk} + 2 \g^{ij} D_\mu X^i D^\mu X^l X^{lj} \nn \\
&\quad + \g^{ij} D_\mu X^8 D^\mu X^8 X^{ij} ) \nn \\
\g^{IJK} X^{ILK} D_\mu X^L D^\mu X^J &\rightarrow  g_{YM} (\g^{ijk} X^{jk} D_\mu X^8 D^\mu X^i + 2 \g^{ij} X^{ik} D_\mu X^k D^\mu X^j \nn \\
&\quad + \g^{ij} X^{ij} D_\mu X^8 D^\mu X^8 ) \nn \\
\g^{\mu \nu} \g^{IJKLM} D_\mu X^I D_\nu X^J X^{KLM} &\rightarrow 3 g_{YM}\g^{\mu \nu} \g^{ijkl} D_\mu X^i D_\nu X^j X^{kl} \nn .
\end{align}
\subsubsection{Three Derivative}
\begin{align}
\g_\mu \g^I D_\nu X^J D^\nu X^J D^\mu X^I &\rightarrow  \g_\mu \g^i D_\nu X^j D^\nu X^j D^\mu X^i + \g_\mu \g^i D_\nu X^8 D^\nu X^8 D^\mu X^i \nn \\
& \quad + \g_\mu \g^8 D_\nu X^j D^\nu X^j D^\mu X^8 + \g_\mu \g^8 D_\nu X^8 D^\nu X^8 D^\mu X^8 \nn \\
\g_\mu \g^I D^\mu X^J D_\nu X^J D^\nu X^I &\rightarrow \g_\mu \g^i D^\mu X^j D_\nu X^j D^\nu X^i + \g_\mu \g^i D^\mu X^8 D_\nu X^8 D^\nu X^i \nn \\
& \quad + \g_\mu D^\mu X^j D_\nu X^j D^\nu X^8 + \g_\mu D^\mu X^8 D_\nu X^8 D^\nu X^8 \nn  \\
\e^{\mu \nu \rho} \g^{IJK} D_\mu X^I D_\nu X^J D_\rho X^K &\rightarrow \e^{\mu \nu \rho} \g^{ijk} D_\mu X^i D_\nu X^j D_\rho X^k + 3 \e^{\mu \nu \rho} \g^{ij} D_\mu X^i D_\nu X^j D_\rho X^8. \nn
\end{align}

\subsection{$\delta X_i$}
\subsubsection{Zero Derivative}
\begin{align}
\be \g^J \psi X^{IKL} X^{JKL} &\ra 2 g^2_{YM}\be \g^j \psi X^{ik} X^{jk} \nn \\
\be \g^{JKL} \psi X^{IJN} X^{KLN} &\ra g^2_{YM}\be \g^{jkl} \psi X^{ij} X^{kl} \nn \\
\be \g^{IJKLM} \psi X^{JKN} X^{LMN} &\ra g^{2}_{YM} \be \g^{ijklm} \psi X^{jk} X^{lm} \nn \\
\end{align}
\subsubsection{One Derivative}
\begin{align}
\be \g_\mu \g^{JKL} \psi D^\mu X^I X^{JKL} &\ra 3g_{YM} \be \g_\mu \g^{jk} \psi D^\mu X^i X^{jk} \nn \\
\be \g_\mu \g^{IKL} \psi D^\mu X^J X^{JKL} &\ra 2g_{YM} \be \g_\mu \g^{ij} \psi D^\mu X^k X^{kj} + g_{YM}\be \g_\mu \g^{ijk} \psi D^\mu X^8 X^{jk} \nn \\
\be \g_\mu \g^K \psi D_\mu X^J X^{JKI} &\ra g_{YM}\be \g_\mu \g^8 \psi D_\mu X^j X^{ij} + g_{YM}\be \g_\mu \g^k \psi D_\mu X^8 X^{ki} \nn \\
\be \g^\mu \g^{IJKLM} \psi D_\mu X^J X^{KLM} &\ra 3g_{YM} \be\g^\mu \g^{ijkl} \psi D_\mu X^j X^{kl} \nn \\
\be \g_\mu \g^{JKL} \psi D^\mu X^K X^{IJL} &\ra 2 g_{YM} \be \g_\mu \g^{jk} \psi D^\mu X^k X^{ij} \nn
\end{align}
\subsubsection{Two Derivative}
\begin{align}
\be \g^I \psi D_\mu X^J D^\mu X^J &\ra \be \g^i \psi D_\mu X^8 D^\mu X^8 + \be \g^i \psi D_\mu X^j D^\mu X^j \nn \\
\be \g^I \g^{\mu \nu} \psi D_\mu X^J D_\nu X^J &\ra  0\nn \\
\be \g^{\mu \nu}\g^{IJK} \psi D_\mu X^J D_\nu X^K &\ra  2\be \g^{\mu \nu} \g^{ij} \psi D_\mu X^j D_\nu X^8 + \be \g^{\mu \nu} \g^{ijk} \psi D_\mu X^j D_\nu X^k \nn \\
\be \g^J \psi D_\mu X^I D^\mu X^J &\ra \be \g^8 \psi D_\mu X^i D^\mu X^8 + \be \g^j \psi D_\mu X^i D^\mu X^j \nn \\
\be \g^J \g^{\mu \nu} \psi D_\mu X^I D_\nu X^J &\ra \be \g^8 \g^{\mu \nu} \psi D_\mu X^i D_\nu X^8 + \be \g^j \g^{\mu \nu} \psi D_\mu X^i D_\nu X^j \nn \\
\be \g^{IJK} \psi D_\mu X^J D^\mu X^K &\ra 0
\end{align}
\subsection{$\delta A_\mu$}
\subsubsection{Zero Derivative}
\begin{align}
\be \g_\mu \chi X^{IJK} X^{IJK} &\ra 3 g^2_{YM} \be \g_\mu \chi X^{ij} X^{ij} \nn \\
\be \g_\mu \g^{IJ} \chi X^{ILM} X^{JLM} &\ra 2 g^2_{YM} \be \g_\mu \g^{ij} \chi X^{il} X^{jl} \nn \\
\be \g_\mu \g_{IJKL} \chi X^{IJN} X^{KLN} &\ra  g^2_{YM}\be \g_\mu \g^{ijkl} \chi X^{ij} X^{kl} \nn \\
\be \g_\mu  \g^{IJKLMN} \chi X^{IJK} X^{LMN} &\ra 0 \nn 
\end{align}
\subsubsection{One Derivative}
\begin{align}
\be \g^{JK} \chi D_\mu X^I X^{IJK} &\ra g_{YM}(\be \g^{jk} \chi D_\mu X^8 X^{jk} + 2 \be \g^j \chi D_\mu X^i X^{ij} )\nn \\
\be \g^{IJKL} \chi D_\mu X^I X^{JKL} &\ra 3 g_{YM} \be \g^{ijk} \chi D_\mu X^i X^{jk} \nn \\
\be \g_{\mu \nu} \g^{JK} D^\nu X^I X^{IJK} &\ra g_{YM}(\be \g_{\mu \nu} \g^{jk} \chi D^\nu X^8 X^{jk} + 2 \be \g_{\mu \nu} \g^j \chi D^\nu X^i X^{ij}) \nn \\
\be \g_{\mu \nu} \g^{IJKL} D^\nu X^I X^{JKL} &\ra 3 g_{YM }\be \g_{\mu \nu} \g^{ijk} \chi D_\mu X^i X^{jk} \nn 
\end{align}
\subsubsection{Two Derivative}
\begin{align}
\be \g_\mu \chi D_\nu X^K D^\nu X_K &\ra \be \g_\mu \chi D_\nu X^8 D^\nu X^8 + \be \g_\mu \chi D_\nu X^i D^\nu X^i \nn \\
\be \g_\mu \g^{IJ} \chi D_\nu X^I D^\nu X^J &\ra \be \g_\mu \g^{ij} \chi D^\nu X^i D^\nu X^j + \be \g_\mu \g^i \chi D_\nu X^i D^\nu X^8 - \be \g_\mu \g^i \chi D_\nu X^8 D^\nu X^i =0\nn \\
\be \g_\nu \chi D^\nu X^K D_\mu X^K &\ra \be \g_\nu \chi D^\nu X^i D_\mu X^i + \be \g_\nu \chi D^\nu X^8 D_\mu X^8 \nn \\
\be \g_\nu \g^{IJ} \chi D^\nu X^I D_\mu X^J &\ra \be \g_\nu \g^{ij} \chi D^\nu X^i D_\nu X^j + \be \g_\nu \g^i \chi D^\nu X^i D_\mu X^8 - \be \g_\nu \g^i \chi D^\nu X^8 D_\mu X^i \nn \\
\be \g_{\mu \nu \lambda} \chi D^\nu X^J D^\lambda X^J &\ra 0 \nn \\
\be \g_{\mu \nu \lambda} \g^{IJ} D^\nu X^I D^\lambda X^J &\ra \be \g_{\mu \nu \lambda} \g^{ij} D^\nu X^i D^\lambda X^j + \be \g_{\mu \nu \lambda} \g^i D^\nu X^i D^\lambda X^8 - \be \g_{\mu \nu \lambda} \g^i D^\nu X^8 D^\lambda X^i \nn 
\end{align}
\end{appendix}
\newpage

\bibliographystyle{JHEP}
\bibliography{DansBib}
\end{document}

%% file: DansDefs.tex
\newcommand{\nn}{\nonumber}

\newcommand{\Tr}{\textrm{Tr}}
\newcommand{\Str}{\textrm{STr}}

\newcommand{\D}{\tilde{D}}

\newcommand{\ra}{\rightarrow}

\newcommand{\bp}{\bar{\psi}}
\newcommand{\e}{\epsilon}
\newcommand{\be}{\bar{\epsilon}}
\newcommand{\g}{\Gamma}